\documentclass{aa}
\def\ref{\bibitem{}}
\usepackage{graphicx}

\begin{document}

\title{Statistics of X-ray observables for the cooling-core and 
non-cooling core galaxy clusters}
\author{Y. Chen\inst{1,2} \and T.H.~Reiprich\inst{3} \and H.~B\"ohringer\inst{2}
                          \and Y.~Ikebe\inst{4} \and Y.-Y.~Zhang\inst{2}}
\institute {
Key Laboratory of Particle Astrophysics,
 Institute of High Energy Physics,
 Chinese Academy of Sciences,
 Beijing 100049, P.R. China
\and
Max-Planck-Institut f\"ur Extraterrestrische Physik, D-85748 Garching,
Germany
\and
Argelander-Institut f\"ur Astronomie, Universit\"at Bonn,
Auf dem H\"ugel 71, 53121 Bonn, Germany
\and
National Museum of Emerging Science and Innovation, Tokyo, 135-0064, Japan}
\offprints{Yong Chen,
\email{ychen@mail.ihep.ac.cn}}
\date{Received ?; accepted ?}

\abstract{
We present a statistical study of the occurrence and effects of the
cooling cores in the clusters of
galaxies in a flux-limited sample, HIFLUGCS, based on
ROSAT and ASCA observations.
About 49\% of the clusters in this sample have a significant,
classically-calculated cooling-flow, mass-deposition rate. The upper
envelope of the derived mass-deposition rate is roughly proportional
to the cluster mass, and the fraction of cooling core clusters is
found to decrease with it. 
The cooling core clusters are found to
have smaller core radii than non-cooling core clusters, while some
non-cooling core clusters have high $\beta$ values ($> 0.8$).
In the relation of the X-ray luminosity vs.
the temperature and the mass, the cooling core
clusters show a significantly higher normalization. A systematic
correlation analysis, also involving relations of the 
gas mass and the total infrared
luminosity, indicates that this bias is shown to be
mostly due to an enhanced X-ray luminosity
for cooling core clusters, while the other parameters, like temperature, mass,
and gas mass may be less affected by the occurrence of a cooling core.
These results may be explained by at least some of the
non-cooling core clusters being in dynamically young states compared with
cooling core clusters, and they may turn into cooling core
clusters in a later evolutionary stage.
}
\titlerunning{Statistics of X-ray observables of galaxy clusters}
\maketitle

\section{Introduction}
Clusters of galaxies are interesting large-scale astrophysical
laboratories offering ideal probes for studying the large-scale
structure of the Universe and for testing the cosmological models
(e.g. Voit 2005). A very important scaling parameter in these studies
is the cluster mass, which cannot easily
be measured unless detailed observations are available.
It is therefore estimated by means of other suitable, easily
obtained global observables such as X-ray luminosity or
X-ray temperature (e.g. Reiprich \& B\"ohringer 2002; Markevitch
1998; Ikebe et al. 2002; Finoguenov et al. 2001; Arnaud et al.
2005). Since the early days of X-ray imaging with the EINSTEIN
satellite, it is apparent that there may be two, 
to some extent distinct, classes of galaxy clusters: clusters with very dense
gaseous core regions, so-called cooling cores, and another type
with shallower cores often exhibiting a more internal structure
(e.g. Jones \& Forman 1984; Ota \& Mitsuda 2004; Peres et al.
1998; Schuecker et al. 2001a). 
In the present paper we explore first the influence of this
dichotomy on the scaling relations between global cluster X-ray
observables and then between the observables and the cluster mass in
order to improve our understanding of how to use these scaling
relations in cosmological applications.

Clusters with dense gaseous cores, which have central cooling times significantly lower than a
Hubble time, have formerly been termed cooling flow clusters, and it was believed that
the intracluster medium (ICM) in these regions cools and condenses, as it
is difficult to avoid cooling in the absence of a very fine-tuned heating 
mechanism (Fabian 1994). A different point of view
not requiring a cooling flow has also been put forward 
based on ASCA spectroscopic results 
(e.g., Ikebe et al. 1999; Makishima et al. 2001, and references therein).
With XMM-Newton observations, it was found that the spectral features predicted by the classical
cooling flow model are not observed in the X-ray spectra of cooling flow regions (e.g. Peterson
et al. 2001, 2003). While a slight temperature decrease by factors up to 2 - 3 towards the
center in cooling cores is observed, the expected features for further cooling are absent.
With high-resolution Chandra observations, a possible fine-tuned heat source has been found
in the interaction of central AGN with the cluster ICM, which is now taken as the
most probable reason for the prevention of massive cooling flows (e.g. David et al. 2001;
B\"ohringer et al. 2002; Fabian et al. 2003; McNamara et al. 2005). Therefore we follow the 
now widely-used
convention to call the clusters in our sample cooling core clusters (CCC) and non-cooling
core clusters (NCCC).

An influence of the CCC or NCCC nature of the clusters on the scaling relations of global
properties has previously been realized, e.g., in the luminosity temperature relation
(e.g. Fabian et al. 1994; Markevitch 1998; McCarthy et al. 2004)
and other parameters (O'Hara et al. 2006). Here
we extend the studies of the influence of CCCs on the scaling relation to the largest X-ray
flux limited sample of galaxy cluster with detailed X-ray data that allow a mass
determination, the HIFLUGCS (the HIghest X-ray FLUx Galaxy Cluster Sample;
Reiprich 2001; Reiprich \& B\"ohringer \cite{Reiprich}). This cluster sample is selected only 
by X-ray flux,
irrespective of the cluster morphology, and we do not know of any signature
of incompleteness in the sample. Therefore it should provide a representative 
mix of cluster morphologies for a given X-ray luminosity, providing the 
correct statistics be applied
to the typical cosmological X-ray survey cluster samples.

In particular we study the segregation of CC and NCC clusters in the $L_X - T$,
$L_X - M$, $M - T$, and the $f_{gas} - T$ relations. A major goal in this study
is to better understand the scatter in these relations, which has to be folded into
the test of large-scale structure measures and cosmological models
(e.g., Ikebe et al. 2002; Stanek et al. 2006). It is especially interesting in the context of the
$L_X - M$ relation given by Reiprich \& B\"ohringer (2002) where the observed scatter is very
large and partly due to the large uncertainties in mass determination. Therefore it
was very difficult to separate the intrinsic scatter from the scatter introduced
by the formal and systematic measurement errors. It is the intrinsic scatter, 
however, that is important
for the application. A difference in the relation amplitude between CCCs and NCCCs
could in principle provide a lower limit to the intrinsic scatter in the $L_X - M$ relation of
Reiprich \& B\"ohringer (2002), if the systematic uncertainties are well controlled,
and thus help to understand the origin
of the scatter better. This is interesting because the best-fit cosmological
parameter values from the WMAP 3rd year data (Spergel et al. 2006) 
applied to compare the predicted and observationally derived HIFLUGCS
cluster mass function provide an indication that the
intrinsic scatter is probably smaller than the systematic measurement
errors that go into the derived mass and X-ray luminosity relation
(Reiprich 2006).
Our study has a lot in common with the work of O'Hara et al. (2006),
but was started independently a few years ago, so we discuss the correspondence
of the two studies throughout the paper.

The paper is organized as follows. In Sect.2 we briefly introduce the sample. In Sect.3,
we present the method of data reduction. We compare the properties of the CCC and NCCC in
Sect.4 and discuss the implications of the results. Section 5 provides a summary.
In the following we adopt a cosmological model with $H_{0}=50$ km s$^{-1}$ Mpc$^{-1}$,
$\Omega_m$=1 and $\Omega_\Lambda$=0, a choice which was mostly made for easier
comparison with previous results.

\section{The sample}
The extended HIFLUGCS sample with 106 clusters and groups of galaxies is used
for the present study. Ninety-two of these clusters have known temperatures determined from
X-ray spectroscopy. Here we use two cluster temperatures:
$T_m$, the emission measure weighted temperature, which is mainly derived
from a single temperature fit to the global X-ray spectrum of the clusters 
(Markevitch et al. 1998; Reiprich 2001 and references therein); and
$T_h$, which is the hotter bulk component of a two-temperature model fitted to
the spectrum (88 of them from ASCA, Ikebe et al. 2002). The $T_h$ was determined
by accepting a small, second lower-temperature component, 
to allow for a low temperature
phase in a possible cooling core in the central cluster region. 
The typical temperature of this second component was about a factor of 2 lower
than the bulk temperature (Ikebe et al. 2002). 
The second component of the two-temperature fit, which is
generally only needed for the cooling core clusters, has a small normalization and
is expected to account for the lower central temperature phase in the cooling cores.
For the clusters with no measured $T_m$ or $T_h$, we derived them using
the $L_X$--$T$ relation of Markevitch (1998) with the $L_X(<2$Mpc) 
uncorrected and $T$ corrected for cooling flows.
The hotter component, $T_h$, which is usually slightly higher than $T_m$,
is expected to provide a good measure of the gravitational
potential depth and the total mass of the clusters.

The X-ray surface brightness profiles are
derived from ROSAT PSPC observations, 36 of them are from RASS observations
(allowing for a large enough field-of-view for the prominent nearby clusters), 
and 70 from pointing observations (Reiprich \& B\"ohringer 2002). 
The large FOV of the ROSAT PSPC allows
us to cover most of the clusters out to $r_{500}$, the radius
at which the mean density of the cluster is 500 times that of a critical
density universe. In addition,
it is worth noting that this flux limited sample has the largest sky area
so far.

The basic properties of the clusters in the HIFLUGCS sample from 
Reiprich \& B\"ohringer (\cite{Reiprich})
are given in Tables 1 and 2. Note that $T_m$, the emission-measure weighted temperature, 
is used in the calculation of some quantities, such as $n_{center}$, $t_{cool}$, 
and $\dot{M}$ in the next section.
In Fig. 1 we show the distribution of the masses and cooling flow mass deposition rates
determined as described in the following sections. The NCC clusters with no significant
mass-deposition rates and with very small or no cooling radii are shown with their mass
distribution at the bottom of the plot. A striking feature of the plot is the ridge of
cooling core clusters with formal mass deposition rates that increase almost linearly
with cluster mass. This leads us to define the class of pronounced cooling core clusters
by a lower limit to the ratio of the formal mass deposition rate to the cluster mass, $M_{500}$.
The ratio value chosen for $\dot M / M_{500}$ of $10^{-13}$ yr$^{-1}$ is indicated in
Fig. 1. A further limit in the mass deposition rate at 0.01 M$_{\odot}$  yr$^{-1}$ is used
to separate small-to-moderate cooling cores from NCCC. The total sample thus splits up into 36
pronounced CCCs, 16 small-to-moderate CCCs, and 54 NCCCs. We make use of this
classification below.

\begin{figure}
\resizebox{\hsize}{!}{\includegraphics[height=7.0cm,width=10.0cm]
{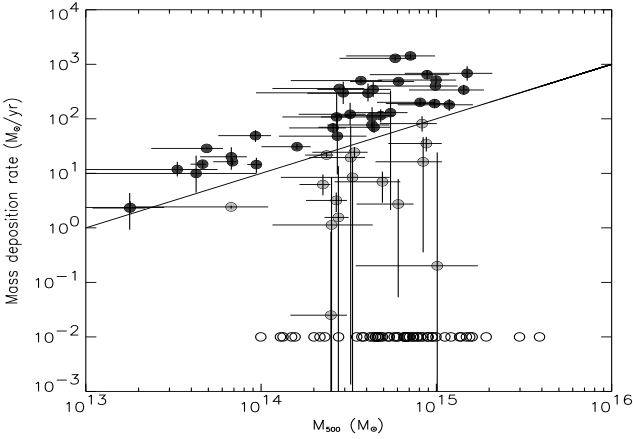}}
\caption{Formally-deduced mass-deposition rates in the frame of
the classical cooling flow model as a function of the total
cluster mass, $M_{500}$. There is a pronounced ridge line of
stronger cooling core clusters with a formal mass deposition rate
almost proportional to the cluster mass. The clusters with
insignificant mass deposition rates below a value of 0.01
M$_{\odot}$  yr$^{-1}$ are plotted according to their total mass
at the bottom of the plot with a formal value of 0.01 M$_{\odot}$
yr$^{-1}$. The line at  $\dot M / M_{500} = 10^{-13}$ yr$^{-1}$
separates strong CCCs from small-to-moderate CCCs. In all figures
of this paper, the filled black circles represent the pronounced CCCs. 
The filled grey circles are the small-to-moderate CCCs and the open
circles are NCCCs. Throughout this paper the error bars are at a
$\pm$68\% confidence level except for $T_m$ of some clusters, $T_h$,
and $L_X$, which are at a 90\% confidence level (and thus a conservative
error estimate).}
\label{ychen:fig1}
\end{figure}

\section{Data analysis}
\subsection{X-ray surface brightness and mass profiles}

To determine the X-ray surface brightness distribution we produced images in
the 0.5 to 2 keV band (PSPC PI channel 52 to 201) and the 
corresponding vignetting-corrected exposure maps. The contaminating point sources 
and obvious substructure
were excised. A center position was obtained from an iterative determination
of the ``center-of-mass'' of the photon distribution in a 7.5 arcmin radius
aperture. The surface brightness profile was then constructed by azimuthal
averaging in concentric bins. This procedure is the same as used in Reiprich
\& B\"ohringer (2002).

We fit the surface brightness profile with a single $\beta$ model (Cavaliere \& Fusco-Femiano \cite{Cavaliere})
\begin{equation}
S(r)=S_0\Big(1+(r/r_c)^2\Big)^{-3\beta+1/2} ,
\end{equation}
where $S_0$ is central brightness (counts/s/pixel$^2$; 1$\arcmin=120$ pixels
for ROSAT PSPC) and $r_c$ the core radius (kpc).
We also try a double $\beta$ model
\begin{equation}
S(r)=S_{01}\alpha_1^{-3\beta_1+1/2}+S_{02}\alpha_2^{-3\beta_2+1/2}
,
\end{equation}

\begin{equation}
\alpha_1=1+(r/r_{c1})^2 ,
\end{equation}

\begin{equation}
\alpha_2=1+(r/r_{c2})^2 ,
\end{equation}
in which the fits of 49 clusters have significantly improved
reduced $\chi^2$ values compared to the fits using a single
$\beta$ model (Table 3). We thus use the
double $\beta$ model for these clusters. Assuming that the
temperature is homogeneous in the cluster, we can calculate the gas
number-density profile $n(r)$. The errors introduced by this
simplification in the presence of temperature variations is only
on the order a few percent, which justifies this approximation.
Assuming, moreover, that the gas is in hydrostatic equilibrium, the
total mass of the cluster can be calculated for a single $\beta $
model as
\begin{equation}
M(r)=\frac{3\beta kT_h r}{G\mu m_p}\frac{(r/r_c)^2}{1+(r/r_c)^2} ,
\end{equation}
where $k$ is the Boltzmann constant, $G$ the gravitational constant,
$\mu$ the molecular weight ($\mu$=0.61), and $m_p$ the proton mass.
For a double $\beta$ model we find
\begin{equation}
M(r)=\frac{3 kT_hr^3}{G\mu m_p}\frac{n_{01}^2 \beta_1\alpha_1^{-3\beta_1-1}/r_{c1}^2
     +n_{02}^2 \beta_2\alpha_2^{-3\beta_2-1}/r_{c2}^2}
     {n_{01}^2\alpha_1^{-3\beta_1}+n_{02}^2\alpha_2^{-3\beta_2}} ,
\end{equation}
where $n_{01}$ and $n_{02}$ are the central equivalent electron number density 
calculated from the two surface brightness components. The central 
electron number density, $n_0$, can be derived from
\begin{equation}
n_0^2=n_{01}^2+n_{02}^2.
\end{equation}

\subsection{Cooling core properties}

The cooling time of the gas is calculated by
\begin{equation}
t_{cool}=\frac{5}{2} \frac{n_e+n_i}{n_e} \frac{kT_m}{n_H \Lambda(A,T_m)} ,
\end{equation}
where $\Lambda(A,T_m)$ is the cooling function of the gas, and $n_e$, $n_i$, and $n_H$
are the number densities of the electrons, ions, and hydrogen, respectively. 
Here we use the abundance $A=0.3$ for all clusters. Note that
for the nearly fully ionized plasma in clusters, $n_e=1.2n_H$ and
$n_i=1.1n_H$ .

Following the previously most frequently-used convention, we define the
cooling radius as the radius where the gas cooling time is equal to the
age of the cluster assumed to be close to the Hubble time ($t_{age} \sim 1/H_0
= 13$ Gyr). The physical meaning of the cooling radius within the classical
cooling flow model is that, within the cooling radius, the gas will lose all
of its energy by X-ray emission and is replaced by ambient hot gas from larger
radii in a steady state inflow. We can therefore calculate the energy loss
rate from the integral of the X-ray emission inside the cooling radius
and the mass inflow rate from the enthalpy influx necessary to compensate
for this energy loss. We then account for the energy gain as the inflowing
gas moves down the gravitational potential gradient. Thus,
the total mass deposition rate within the shell $i$ can be determined by
\begin{equation}
\dot M(i)= \frac{n_e(i) n_H(i) \Lambda(A,T_m) V(i)+\frac{5}{2}\frac{kT_m}{\mu m_p}\dot M(i-1)}
{\frac{5}{2}\frac{kT_m}{\mu m_p} + \Phi(i+1)-\Phi(i)} ,
\end{equation}
where $V(i)$ is volume of the shell $i$, and $\Phi(i+1)$ and $\Phi(i)$
are the gravitational potential of the shell $i+1$ and $i$,
respectively. For a single $\beta$ model, $\Phi(i)$ can be
calculated from
\begin{equation}
\Phi(i)=\frac{3}{2} \frac{\beta kT_h}{\mu m_p} \ln{(1+(\frac{r(i)}{r_c})^2)} .
\end{equation}

When we use the double $\beta$ model, these formulae will change accordingly, and we limit
their writing-out for brevity.

\subsection{Error estimate}
We adopted a Monte-Carlo method to estimate the errors of the
cluster properties derived in the previous subsection, such as the
mass and mass deposition rate. In the calculation of the mass of
the clusters, we assumed a polytropic index with a value of
$\gamma=1$. From previous observational studies, the range of
$\gamma$-values is constrained to be between 0.9 and 1.3 (e.g.
Finoguenov \cite{Fino01}; Pratt et al. 2006). The main errors are
from the temperature and the $\gamma$. We used a $\beta\gamma$
model (e.g. Ettori \cite{Ettori00}) to estimate the errors of the
mass deposition rate and assumed a polytropic index greater than
0.9 and following a Gaussian distribution with a mean value of
1.15 and a variance of 0.15 as derived in
Finoguenov et al. (2001). For each cluster in the HIFLUGCS, 
we created
a cluster sample with 1000 clusters with simulated $T_m$, $T_h$, $S$,
$\beta$, $r_c$, and $\gamma$ according to their own errors. We
calculated other properties (e.g. $n$, $t_{cool}$, $\dot M$, $M$
and so on) of each simulated cluster and then obtained the
errors.
\section{Statistical properties}

In the following we investigate the relations between several
observables and the cluster mass. For all the relations we use
the BCES-Bisector fit of Akritas \& Bershady (1996). The fits are
performed with the logarithmic values of the parameters and quoted
in the form
$$ \log_{10}(Y) = A + B \cdot \log_{10}(X)$$
in Table 7. The median X-ray luminosity, $L_X$(0.1 - 2.4 keV), the median
temperature, $T_m$, and the median cluster mass, $M_{500}$ of the sample
are $2.9~ 10^{44}$ erg s$^{-1}$, 4.1 keV, and $4.8~10^{14}$ M$_{\odot}$, respectively.
Therefore we use the values $1~ 10^{44}$ erg s$^{-1}$, 4 keV, and $5~10^{14}$
M$_{\odot}$, respectively, as pivot points for the fits of the relations.

\begin{figure}
\resizebox{\hsize}{!}{\includegraphics[height=7.0cm,width=10.0cm]
{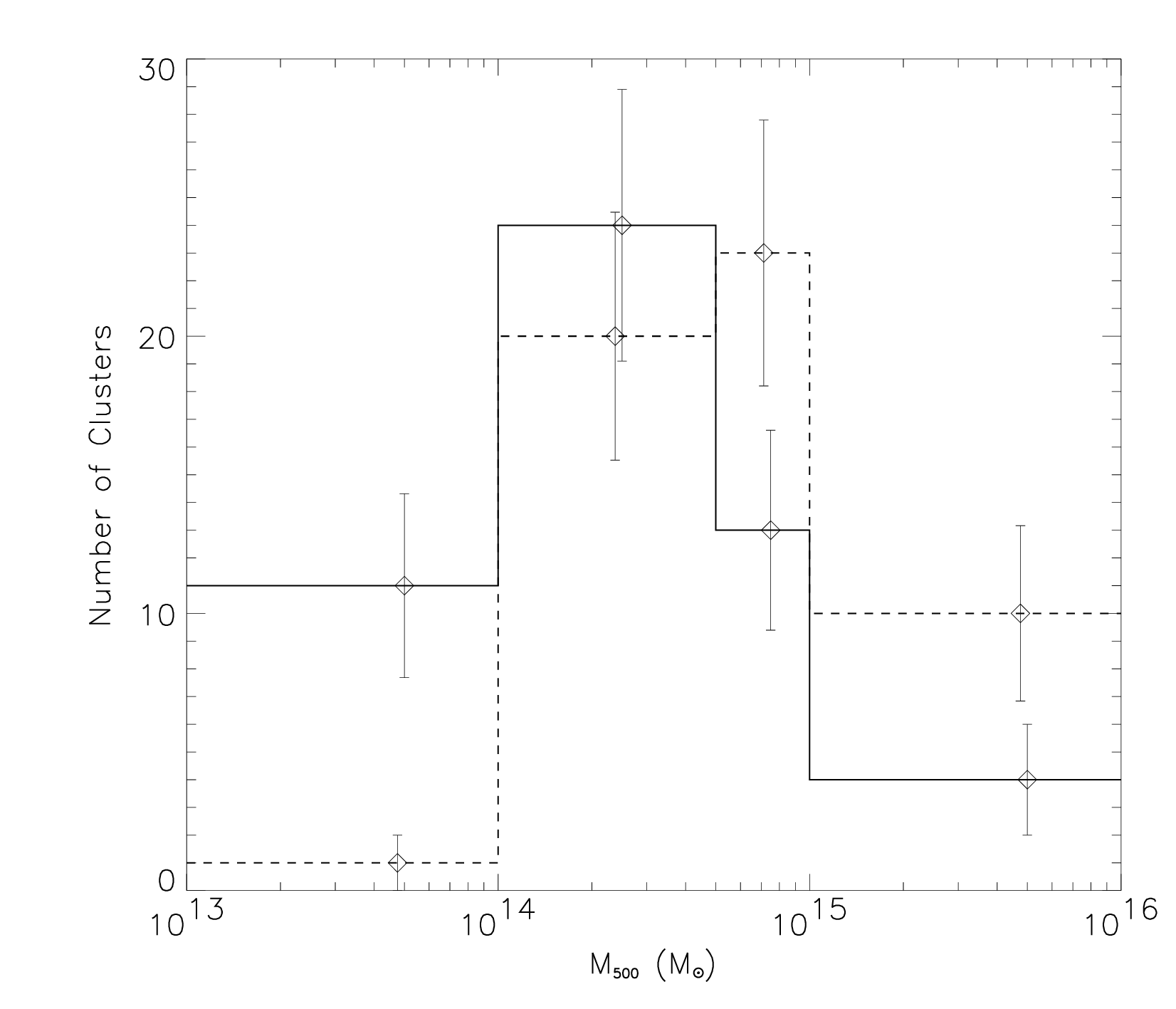}}
\caption{ The numbers of CCCs (solid lines) and NCCCs (dashed lines)
versus the mass of the clusters $M_{500}$. Note that the fraction of NCCCs
increases with $M_{500}$. }
\label{ychen:fig2}
\end{figure}

\subsection{Basic properties}

In Fig. 2 we show the number of CCCs (here including the strong CCC and
moderate CCC) and NCCCs in the sample as a function of the
cluster mass. The fraction of NCCCs clearly increases with $M_{500}$. 
This is also
seen in the smaller flux-limited sample used by O'Hara et al. (2006).
At the low $M_{500}$
end, this may partly be due to some small mass non-cooling core groups
possibly having low luminosities and not reaching the flux limit of HIFLUGCS.
In general, however, the main reasons may be that the fraction of dynamically
young clusters increases with cluster mass and that these clusters do not 
generally feature cooling cores. In addition the ICM is hotter in more massive 
clusters making the
radiative heat loss relatively slower. This is an important statistical
property of the cluster sample to keep in mind, since any segregation of CCCs
and NCCCs in the parameter relations can then also introduce a mass-dependent
effect in the relations of the combined sample.

Figure 3 shows the distribution of the values for the core radius,
$r_c$, and slope parameter $\beta$ from the fit of the
$\beta$-model to the X-ray surface-brightness profiles of the
clusters in the sample. As in previous work (e.g. Jones \&
Forman 1984; White et al. \cite{White}; Ota \& Mitsuda 2004), the
CC clusters segregate very distinctly at lower values of the
core radius than the NCC clusters. Furthermore, we find that the
high the relative $\dot{M_{r}}$, the low the $r_c$. In
addition, it was found that some NCCCs have high $\beta$ values
with $\beta > 0.8$, while such high values are not found among the
CCCs. There are 2 reasons for this behavior. For similar gravitational
potential shapes in CCCs and NCCCs, the CC clusters with central
temperature drop, and a corresponding central ICM density increase
in pressure equilibrium feature X-ray surface brightness cusps
that are fit by smaller core radii (Jones \& Forman 1984). In
addition, NCCCs are often dynamically young, featuring substructure,
elongations, or disturbed core regions which result in inflated
core radii, that in turn lead to steeper outer surface brightness
slopes.

\begin{figure}
\resizebox{\hsize}{!}{\includegraphics[height=7.0cm,width=10.0cm]
{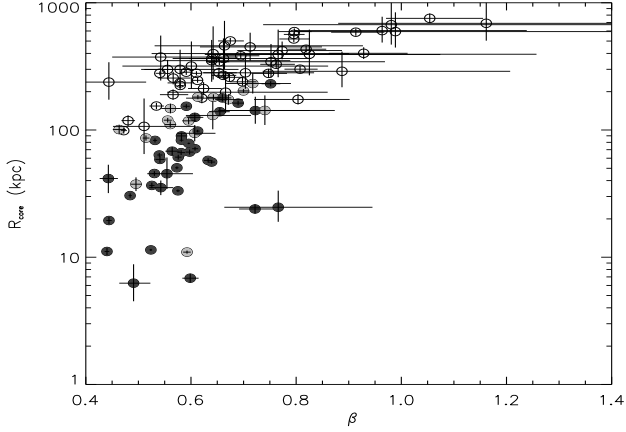}}
\caption{ The $r_c$--$\beta$ diagram, showing the cooling-core clusters 
have smaller core radii. }
\label{ychen:fig3}
\end{figure}

\begin{figure}
\resizebox{\hsize}{!}{\includegraphics[height=7.0cm,width=10.0cm]{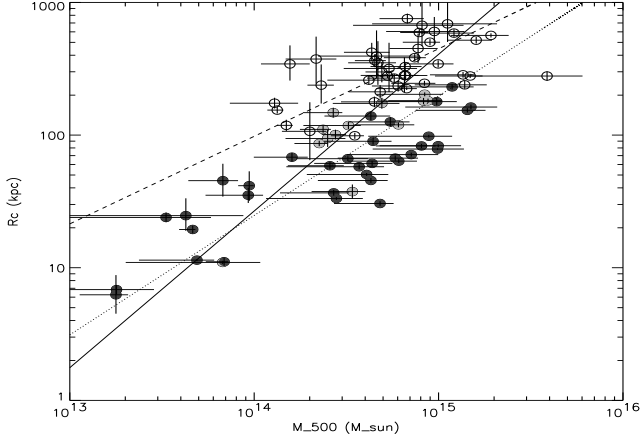}}
\caption{The $r_c$--$M_{500}$ diagram. Here $r_c$ increases with $M_{500}$
faster than the self-similar expectation $r_c \propto M^{1/3}_{500}$.
The solid line represents the BCES bisector fit for all 106 clusters
with the form $ \log_{10}\left({r_{c}
 \over 100  {\rm kpc}}\right) = A
+ B\cdot \log_{10} \left({M_{500} \over 5 10^{14}{\rm
M_{\odot}}}\right)$. The dashed and the dotted lines represent
the fits for the NCCCs and CCCs, respectively (see Table 6). }
\label{ychen:fig4}
\end{figure}

In Fig. 4 we explore the relation of the core radius, $r_c$, with the cluster
mass, $M_{500}$. Assuming that clusters have a strictly self-similar shape, we expect
that any characteristic radius scales as $r \propto M^{1/3}$. The results of
the power-law scaling relation fits to the data are given in Table 6 and shown
in Fig. 4. For both subsamples, CCCs and NCCCs, the observed slope is steeper
than this simple expectation; that is, the core radius increases with mass faster 
than expected. The explanation for this behavior is probably not trivial.
For the NCCCs, the reason might again be that the fraction of dynamically young
clusters with inflated core radii may be larger for higher cluster masses.
For the CCCs, it might be the
increasing dominance of the central cluster galaxy with decreasing cluster mass
that makes the core region relatively more compact for less massive systems.

It is interesting to note that the relation fitted to the complete
sample is steeper than each of the separately fitted relations. This is exactly
the effect mentioned above. It is the result of an offset
in the relation of the two subsamples (significantly smaller core radii for
the CCCs) and, in addition, of a biased
distribution of the clusters in the two subsamples
with more CCCs at the low-mass end and more NCCCs at the high-mass end.
Among all the plots we show in this paper, this is the relation where
this effect is most pronounced.

\begin{figure}
\resizebox{\hsize}{!}{\includegraphics[height=12.0cm,width=14.0cm]{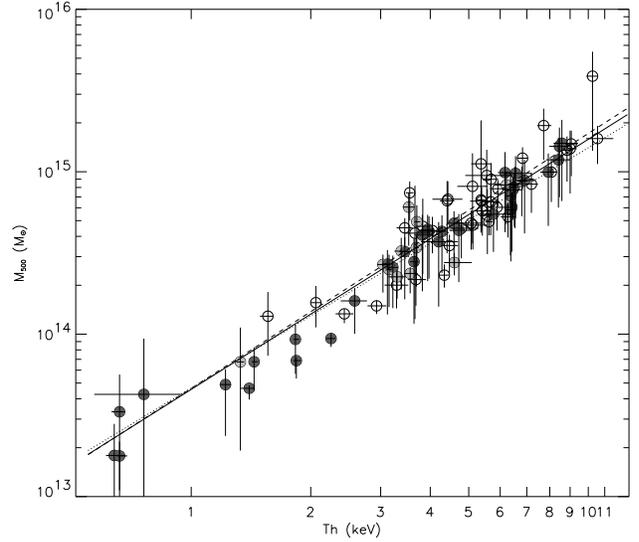}}
\caption{The $M_{500}$--$T_h$ relation. The slopes derived from the total 88 clusters with
measured $T_h$ from ASCA (solid line), CCCs (dotted line), and NCCCs (dashed line) are consistent 
with each other.}
\label{ychen:fig5}
\end{figure}

\begin{figure}
\resizebox{\hsize}{!}{\includegraphics[height=12.0cm,width=14.0cm]{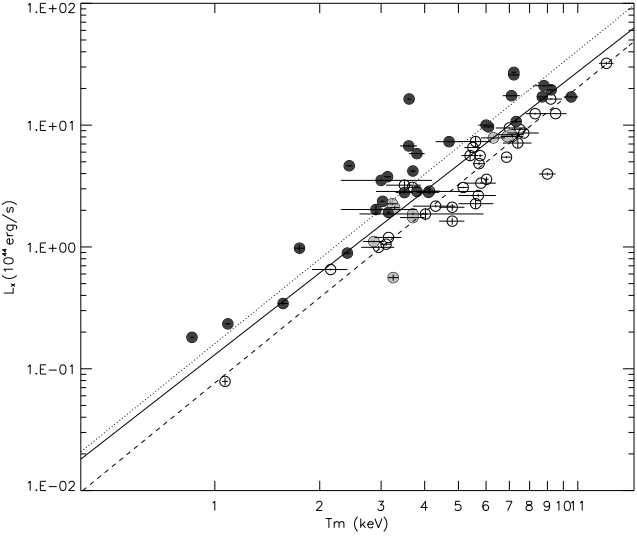}}
\caption{The $L_X$--$T_m$ relation. The solid line shows the power-law relation fit to
all the data, and the dotted and dashed lines are those for the CCCs and NCCCs, 
respectively. $T_m$ with a, b, c, and d in Table 1
are selected to plot here. Note that $T_m$ with d is replaced by the 
temperature measured from the central 2$\arcmin$ or 3$\arcmin$ 
region (see Table 2 in Fukazawa et al. 2000).}
\label{ychen:fig6}
\end{figure}

\begin{figure}
\resizebox{\hsize}{!}{\includegraphics[height=12.0cm,width=14.0cm]{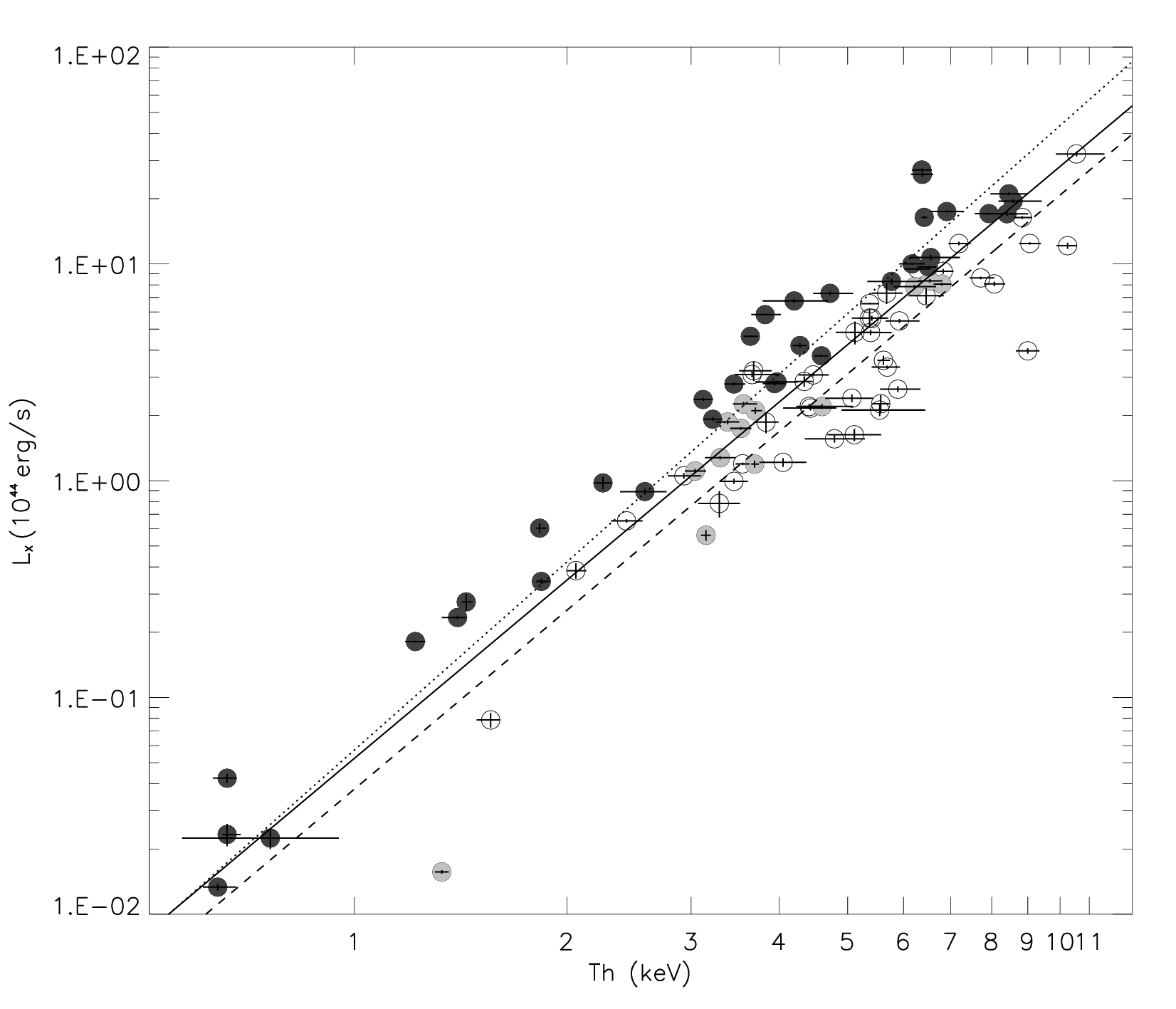}}
\caption{The $L_X$--$T_h$ relation. The solid line shows the power-law relation fit to
all the data, and the dotted and dashed lines are those for the CCCs and NCCCs, respectively. }
\label{ychen:fig7}
\end{figure}

\begin{figure}
\resizebox{\hsize}{!}{\includegraphics[height=12.0cm,width=14.0cm]{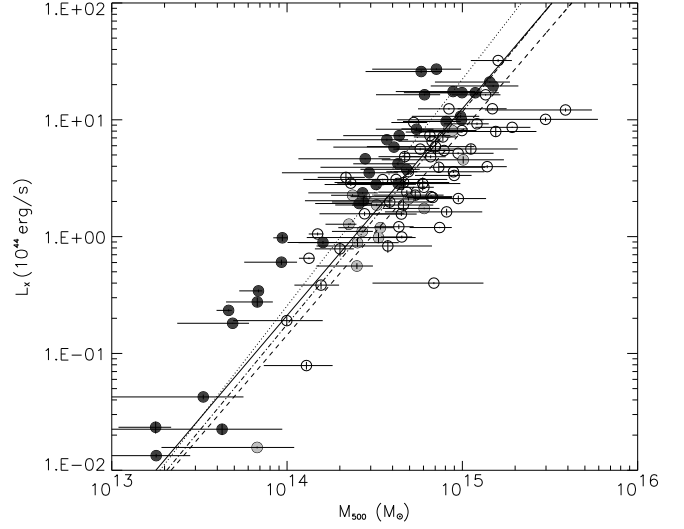}}
\caption{The $L_X$--$M_{500}$ relation. The dot-dashed line shows the power-law 
relation fit to all the data (106 clusters). The solid line shows that for
all the data with ASCA measured $T_h$, 
and the dotted and dashed lines are those for the CCCs and NCCCs, respectively.}
\label{ychen:fig8}
\end{figure}

\begin{figure}
\resizebox{\hsize}{!}{\includegraphics[height=12.0cm,width=14.0cm]{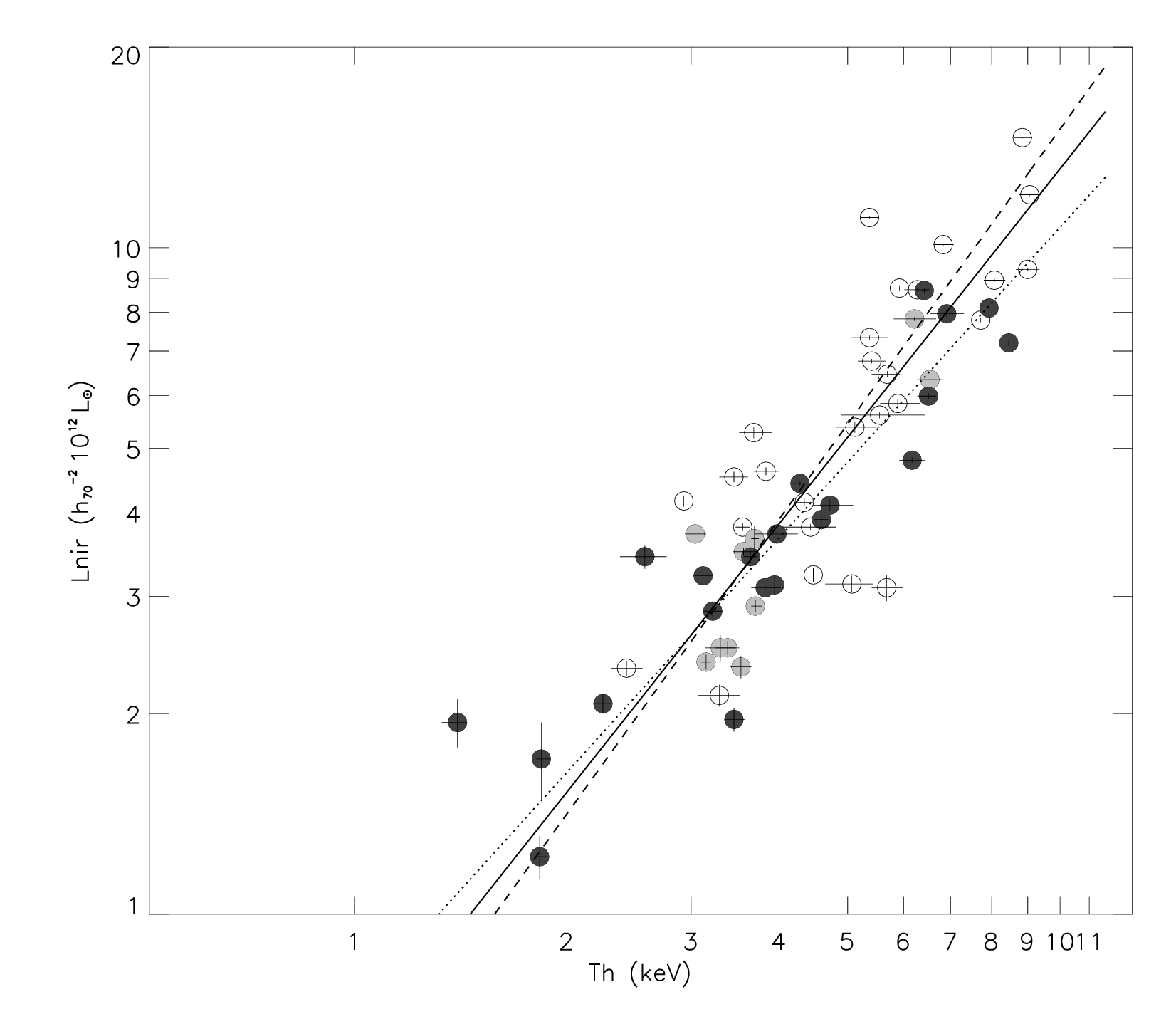}}
\caption{The $L_{nir}$--$T_h$ relation. The solid line shows the 
power-law relation fit to
all the data and the dotted and dashed lines are those for the 
CCCs and NCCCs, respectively. }
\label{ychen:fig9}
\end{figure}

\begin{figure}
\resizebox{\hsize}{!}{\includegraphics[height=12.0cm,width=14.0cm]
{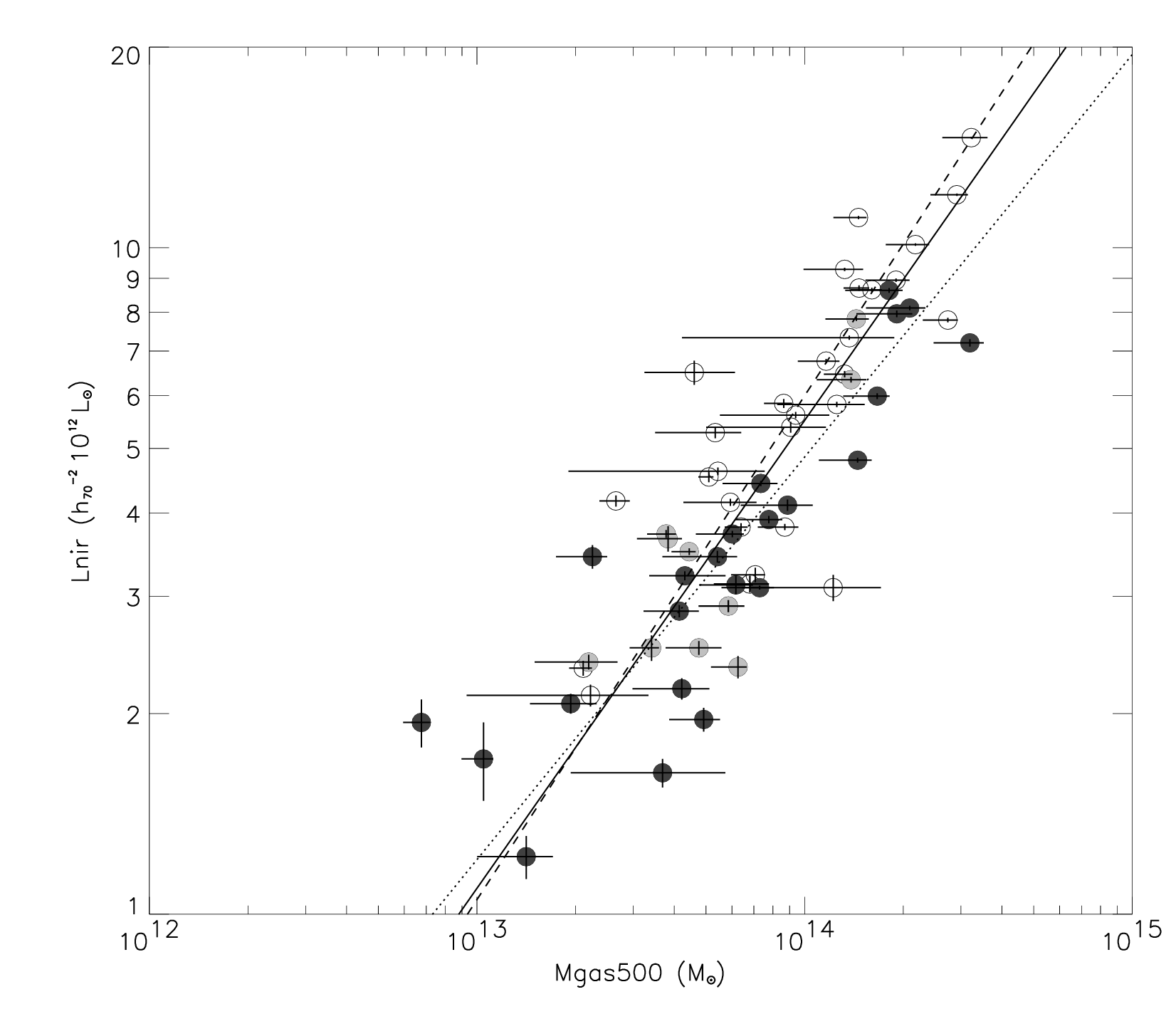}}
\caption{The $L_{nir}-M_{gas,500}$ relation. The solid line shows the 
power-law relation fit to
all the data, and the dotted and dashed lines are those for the
CCCs and NCCCs, respectively. }
\label{ychen:fig10}
\end{figure}

\begin{figure}
\resizebox{\hsize}{!}{\includegraphics[height=12.0cm,width=14.0cm]{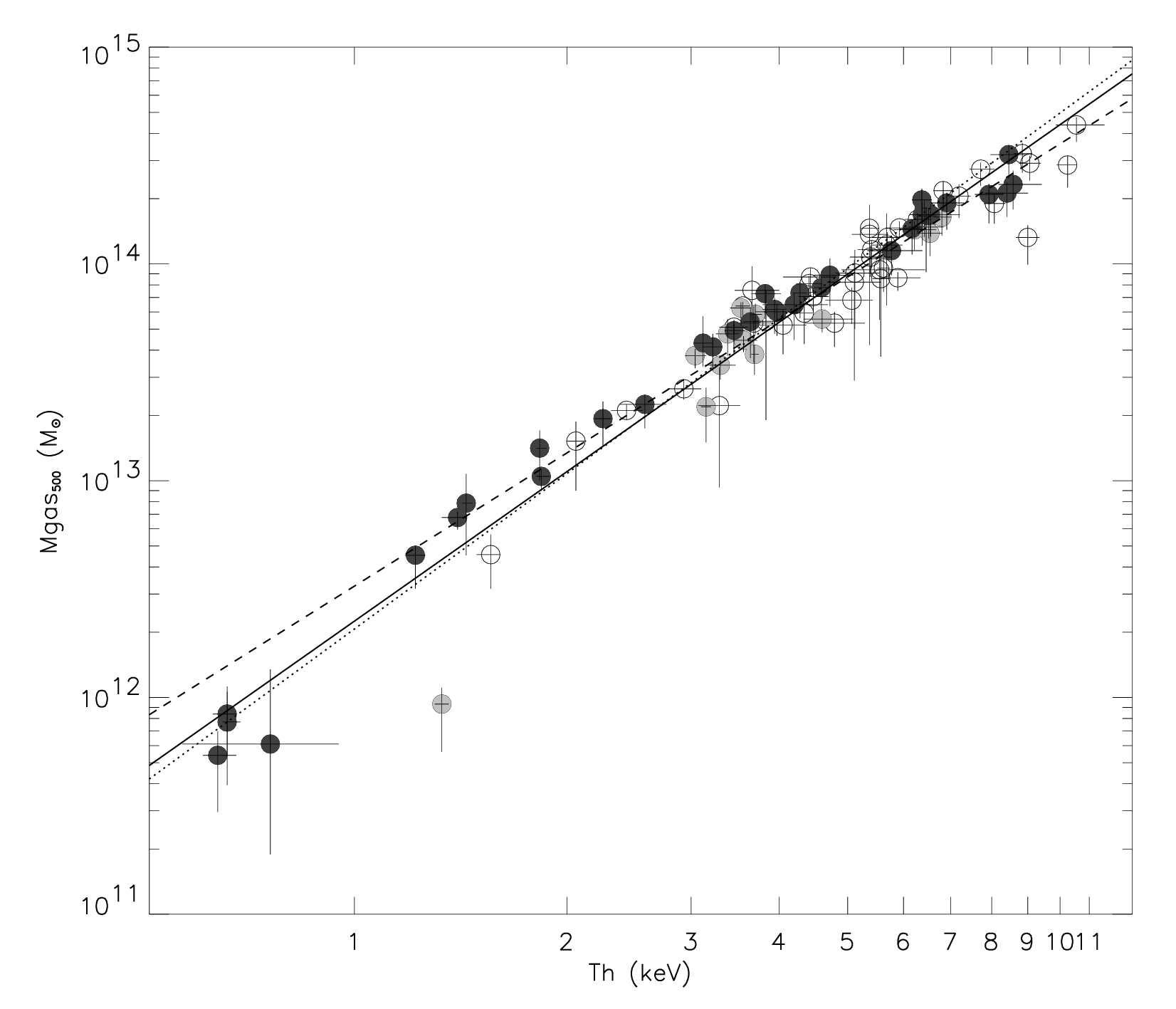}}
\caption{The $M_{gas,500}$--$T_h$ relation. The solid line shows the 
power-law relation fit to
all the data, and the dotted and dashed lines are those for the
CCCs and NCCCs, respectively.}
\label{ychen:fig11}
\end{figure}

\begin{figure}
\resizebox{\hsize}{!}{\includegraphics[height=12.0cm,width=14.0cm]
{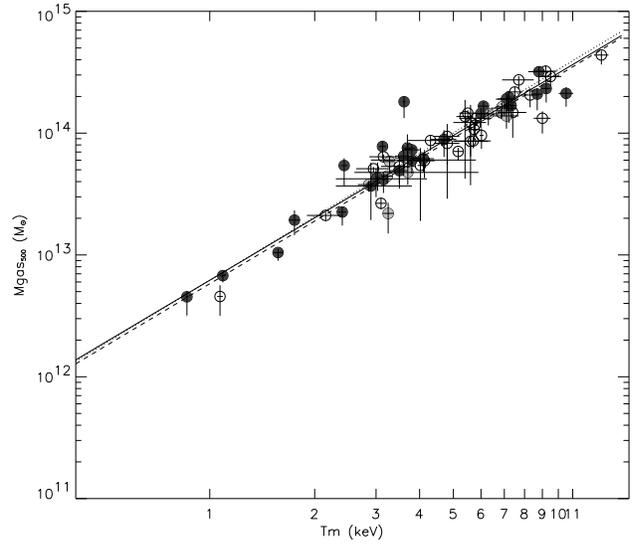}}
\caption{The $M_{gas,500}$--$T_m$ relation. $T_m$ with a, b, c and d in Table 1
are selected to plot here. Note $T_m$ with d is replaced by the 
temperature measured from the central 2$\arcmin$ or 3$\arcmin$ 
region (see Table 2 in Fukazawa et al. 2000).
Symbols have the same meanings as in Fig. 11.}
\label{ychen:fig12}
\end{figure}

\begin{figure}[ht]
\resizebox{\hsize}{!}{\includegraphics[height=12.0cm,width=14.0cm]{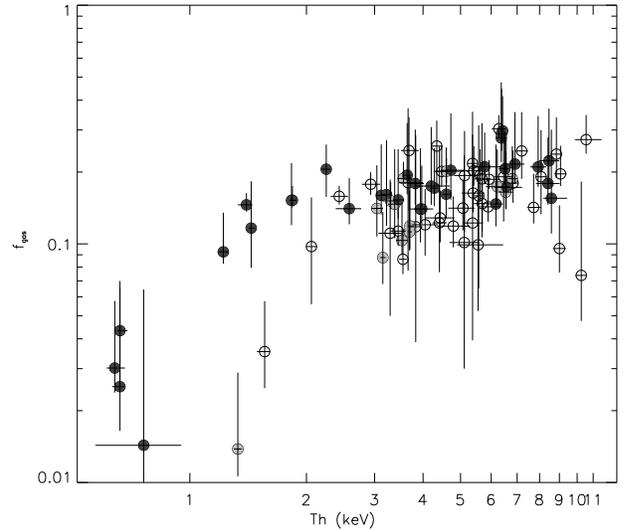}}
\caption{The $f_{gas,500}$--$T_h$ relation. }
\label{ychen:fig13}
\end{figure}

\renewcommand{\arraystretch}{0.65}
\begin{table*}
\caption[]{Cluster properties.}
\vskip -0.2cm
\begin{tabular}{cccccccc}
\hline\noalign{}
 Name  & z   & $T_m$ (keV) & $\beta$ & $r_c$ (kpc) & $n_{center}$ (10$^{-2}$ cm$^{-3}$) & $t_{cool}$ ($10^{10}$ yr) &
 $\dot{M}$ (M$_{\sun}$/yr)    \\
\hline\noalign{}
 2A0335   & 0.0349  &   $^d$3.01$^{+  0.07}_{-  0.07  }$  &   0.575$^{+ 0.004}_{- 0.003  }$  &     33$^{+     0}_{-     0  }$  &   5.47$^{+  0.06}_{-  0.06  }$  &   0.10$^{+  0.00}_{-  0.00  }$  &     360$^{+    20}_{-    17  }$ \\
 A0085    & 0.0556  &   $^a$6.10$^{+  0.20}_{-  0.20  }$  &   0.532$^{+ 0.004}_{- 0.004  }$  &     82$^{+     3}_{-     3  }$  &   2.57$^{+  0.10}_{-  0.10  }$  &   0.33$^{+  0.02}_{-  0.02  }$  &     200$^{+    33}_{-    27  }$ \\
 A0119    & 0.0440  &   $^a$5.80$^{+  0.60}_{-  0.60  }$  &   0.675$^{+ 0.026}_{- 0.023  }$  &    501$^{+    27}_{-    26  }$  &   0.15$^{+  0.01}_{-  0.01  }$  &   5.49$^{+  0.63}_{-  0.58  }$  &       0$^{+     0}_{-     0  }$ \\
 A0133    & 0.0569  &   $^c$3.80$^{+  2.00}_{-  0.90  }$  &   0.530$^{+ 0.004}_{- 0.004  }$  &     45$^{+     1}_{-     1  }$  &   2.80$^{+  0.08}_{-  0.08  }$  &   0.23$^{+  0.09}_{-  0.04  }$  &     108$^{+    55}_{-    51  }$ \\
 A0262    & 0.0161  &   $^d$2.15$^{+  0.06}_{-  0.06  }$  &   0.443$^{+ 0.018}_{- 0.017  }$  &     41$^{+    11}_{-     9  }$  &   0.81$^{+  0.13}_{-  0.09  }$  &   0.56$^{+  0.07}_{-  0.08  }$  &      14$^{+     4}_{-     4  }$ \\
 A0399    & 0.0715  &   $^a$7.40$^{+  0.70}_{-  0.70  }$  &   0.713$^{+ 0.137}_{- 0.095  }$  &    449$^{+   131}_{-    99  }$  &   0.22$^{+  0.04}_{-  0.03  }$  &   4.32$^{+  0.78}_{-  0.70  }$  &       0$^{+     0}_{-     0  }$ \\
 A0400    & 0.0240  &   $^d$2.31$^{+  0.14}_{-  0.14  }$  &   0.534$^{+ 0.014}_{- 0.013  }$  &    154$^{+     9}_{-     8  }$  &   0.20$^{+  0.01}_{-  0.01  }$  &   2.39$^{+  0.13}_{-  0.12  }$  &       0$^{+     0}_{-     0  }$ \\
 A0401    & 0.0748  &   $^a$8.30$^{+  0.50}_{-  0.50  }$  &   0.613$^{+ 0.010}_{- 0.010  }$  &    245$^{+    11}_{-    10  }$  &   0.60$^{+  0.05}_{-  0.04  }$  &   1.67$^{+  0.15}_{-  0.14  }$  &       0$^{+     0}_{-     0  }$ \\
 A0478    & 0.0900  &   $^a$7.10$^{+  0.40}_{-  0.40  }$  &   0.613$^{+ 0.004}_{- 0.004  }$  &     98$^{+     2}_{-     2  }$  &   3.55$^{+  0.15}_{-  0.14  }$  &   0.26$^{+  0.02}_{-  0.02  }$  &     645$^{+   137}_{-   113  }$ \\
 A0496    & 0.0328  &   $^d$4.13$^{+  0.08}_{-  0.08  }$  &   0.484$^{+ 0.003}_{- 0.003  }$  &     30$^{+     1}_{-     1  }$  &   4.07$^{+  0.29}_{-  0.26  }$  &   0.17$^{+  0.01}_{-  0.01  }$  &     114$^{+    35}_{-    28  }$ \\
 A0576    & 0.0381  &   $^b$4.02$^{+  0.07}_{-  0.07  }$  &   0.825$^{+ 0.432}_{- 0.185  }$  &    394$^{+   221}_{-   125  }$  &   0.16$^{+  0.06}_{-  0.04  }$  &   4.27$^{+  1.22}_{-  1.09  }$  &       0$^{+     0}_{-     0  }$ \\
 A0754    & 0.0528  &   $^a$9.00$^{+  0.50}_{-  0.50  }$  &   0.698$^{+ 0.027}_{- 0.024  }$  &    239$^{+    17}_{-    16  }$  &   0.44$^{+  0.02}_{-  0.02  }$  &   2.41$^{+  0.13}_{-  0.13  }$  &       0$^{+     0}_{-     0  }$ \\
 A1060    & 0.0114  &   $^d$3.24$^{+  0.06}_{-  0.06  }$  &   0.607$^{+ 0.040}_{- 0.034  }$  &     94$^{+    15}_{-    12  }$  &   0.47$^{+  0.04}_{-  0.04  }$  &   1.29$^{+  0.11}_{-  0.11  }$  &       0$^{+     0}_{-     0  }$ \\
 A1367    & 0.0216  &   $^d$3.55$^{+  0.08}_{-  0.08  }$  &   0.695$^{+ 0.035}_{- 0.032  }$  &    383$^{+    24}_{-    22  }$  &   0.15$^{+  0.01}_{-  0.01  }$  &   4.29$^{+  0.22}_{-  0.21  }$  &       0$^{+     0}_{-     0  }$ \\
 A1644    & 0.0474  &   4.70$^{+  0.90}_{-  0.70  }$  &   0.579$^{+ 0.111}_{- 0.074  }$  &    299$^{+   127}_{-    92  }$  &   0.28$^{+  0.08}_{-  0.06  }$  &   2.65$^{+  0.96}_{-  0.67  }$  &       0$^{+     0}_{-     0  }$ \\
 A1650    & 0.0845  &   $^a$5.60$^{+  0.60}_{-  0.60  }$  &   0.704$^{+ 0.131}_{- 0.081  }$  &    281$^{+   104}_{-    70  }$  &   0.43$^{+  0.09}_{-  0.06  }$  &   1.87$^{+  0.33}_{-  0.36  }$  &       0$^{+     0}_{-     0  }$ \\
 A1651    & 0.0860  &   $^a$6.30$^{+  0.50}_{-  0.50  }$  &   0.643$^{+ 0.014}_{- 0.013  }$  &    180$^{+     9}_{-     9  }$  &   1.10$^{+  0.07}_{-  0.07  }$  &   0.78$^{+  0.07}_{-  0.07  }$  &      81$^{+    30}_{-    23  }$ \\
 A1736    & 0.0461  &   $^a$3.50$^{+  0.40}_{-  0.40  }$  &   0.542$^{+ 0.147}_{- 0.092  }$  &    374$^{+   177}_{-   129  }$  &   0.13$^{+  0.04}_{-  0.02  }$  &   4.92$^{+  1.27}_{-  1.28  }$  &       0$^{+     0}_{-     0  }$ \\
 A1795    & 0.0616  &   $^a$6.00$^{+  0.30}_{-  0.30  }$  &   0.596$^{+ 0.003}_{- 0.002  }$  &     78$^{+     1}_{-     1  }$  &   2.88$^{+  0.05}_{-  0.05  }$  &   0.29$^{+  0.01}_{-  0.01  }$  &     399$^{+    47}_{-    39  }$ \\
 A2029    & 0.0767  &   $^a$8.70$^{+  0.30}_{-  0.30  }$  &   0.582$^{+ 0.004}_{- 0.004  }$  &     83$^{+     2}_{-     2  }$  &   3.90$^{+  0.15}_{-  0.15  }$  &   0.26$^{+  0.01}_{-  0.01  }$  &     514$^{+    93}_{-    78  }$ \\
 A2052    & 0.0348  &   $^b$3.03$^{+  0.04}_{-  0.04  }$  &   0.526$^{+ 0.005}_{- 0.005  }$  &     36$^{+     1}_{-     1  }$  &   2.85$^{+  0.66}_{-  0.62  }$  &   0.20$^{+  0.06}_{-  0.04  }$  &     108$^{+   188}_{-    49  }$ \\
 A2063    & 0.0354  &   $^d$3.68$^{+  0.11}_{-  0.11  }$  &   0.561$^{+ 0.011}_{- 0.011  }$  &    110$^{+     6}_{-     6  }$  &   0.96$^{+  0.00}_{-  0.00  }$  &   0.67$^{+  0.02}_{-  0.02  }$  &      21$^{+     2}_{-     2  }$ \\
 A2065    & 0.0721  &   $^a$5.40$^{+  0.30}_{-  0.30  }$  &   1.162$^{+ 0.734}_{- 0.282  }$  &    689$^{+   360}_{-   186  }$  &   0.20$^{+  0.07}_{-  0.04  }$  &   3.99$^{+  1.15}_{-  1.04  }$  &       0$^{+     0}_{-     0  }$ \\
 A2142    & 0.0899  &   $^a$8.80$^{+  0.60}_{-  0.60  }$  &   0.591$^{+ 0.006}_{- 0.006  }$  &    153$^{+     5}_{-     5  }$  &   1.61$^{+  0.07}_{-  0.07  }$  &   0.64$^{+  0.04}_{-  0.04  }$  &     337$^{+    82}_{-    61  }$ \\
 A2147    & 0.0351  &   $^d$4.91$^{+  0.28}_{-  0.28  }$  &   0.444$^{+ 0.071}_{- 0.046  }$  &    237$^{+   102}_{-    64  }$  &   0.17$^{+  0.04}_{-  0.03  }$  &   4.41$^{+  0.92}_{-  0.87  }$  &       0$^{+     0}_{-     0  }$ \\
 A2163    & 0.2010  &  $^b$13.29$^{+  0.64}_{-  0.64  }$  &   0.796$^{+ 0.030}_{- 0.028  }$  &    519$^{+    30}_{-    29  }$  &   0.53$^{+  0.02}_{-  0.02  }$  &   2.37$^{+  0.12}_{-  0.11  }$  &       0$^{+     0}_{-     0  }$ \\
 A2199    & 0.0302  &   $^d$4.10$^{+  0.08}_{-  0.08  }$  &   0.655$^{+ 0.019}_{- 0.021  }$  &    139$^{+    10}_{-     9  }$  &   0.83$^{+  0.03}_{-  0.03  }$  &   0.83$^{+  0.03}_{-  0.03  }$  &      77$^{+     7}_{-     6  }$ \\
 A2204    & 0.1523  &   $^b$7.21$^{+  0.25}_{-  0.25  }$  &   0.597$^{+ 0.008}_{- 0.007  }$  &     67$^{+     3}_{-     2  }$  &   5.07$^{+  0.12}_{-  0.11  }$  &   0.18$^{+  0.01}_{-  0.01  }$  &    1287$^{+   122}_{-   129  }$ \\
 A2244    & 0.0970  &   7.10$^{+  5.00}_{-  2.20  }$  &   0.607$^{+ 0.016}_{- 0.015  }$  &    125$^{+    10}_{-    10  }$  &   1.20$^{+  0.06}_{-  0.05  }$  &   0.76$^{+  0.35}_{-  0.18  }$  &     129$^{+   199}_{-   127  }$ \\
 A2255    & 0.0800  &   $^b$6.87$^{+  0.20}_{-  0.20  }$  &   0.797$^{+ 0.033}_{- 0.030  }$  &    593$^{+    35}_{-    32  }$  &   0.18$^{+  0.02}_{-  0.02  }$  &   4.96$^{+  0.45}_{-  0.44  }$  &       0$^{+     0}_{-     0  }$ \\
 A2256    & 0.0601  &   $^a$7.50$^{+  0.40}_{-  0.40  }$  &   0.914$^{+ 0.054}_{- 0.047  }$  &    587$^{+    39}_{-    36  }$  &   0.26$^{+  0.01}_{-  0.01  }$  &   3.62$^{+  0.19}_{-  0.21  }$  &       0$^{+     0}_{-     0  }$ \\
 A2589    & 0.0416  &   $^c$3.70$^{+  2.20}_{-  1.10  }$  &   0.596$^{+ 0.013}_{- 0.012  }$  &    118$^{+     7}_{-     7  }$  &   0.74$^{+  0.07}_{-  0.07  }$  &   0.88$^{+  0.37}_{-  0.24  }$  &      19$^{+    53}_{-    19  }$ \\
 A2597    & 0.0852  &   $^a$3.60$^{+  0.20}_{-  0.20  }$  &   0.633$^{+ 0.008}_{- 0.008  }$  &     57$^{+     2}_{-     2  }$  &   3.63$^{+  0.08}_{-  0.07  }$  &   0.17$^{+  0.01}_{-  0.01  }$  &     501$^{+    58}_{-    51  }$ \\
 A2634    & 0.0312  &   $^d$3.70$^{+  0.28}_{-  0.28  }$  &   0.640$^{+ 0.051}_{- 0.043  }$  &    364$^{+    43}_{-    38  }$  &   0.24$^{+  0.02}_{-  0.02  }$  &   2.70$^{+  0.28}_{-  0.25  }$  &       0$^{+     0}_{-     0  }$ \\
 A2657    & 0.0404  &   $^a$3.70$^{+  0.30}_{-  0.30  }$  &   0.556$^{+ 0.008}_{- 0.007  }$  &    119$^{+     5}_{-     4  }$  &   0.54$^{+  0.00}_{-  0.00  }$  &   1.20$^{+  0.07}_{-  0.07  }$  &       2$^{+     5}_{-     2  }$ \\
 A3112    & 0.0750  &   $^a$4.70$^{+  0.40}_{-  0.40  }$  &   0.576$^{+ 0.006}_{- 0.006  }$  &     61$^{+     2}_{-     2  }$  &   3.52$^{+  0.27}_{-  0.23  }$  &   0.21$^{+  0.02}_{-  0.02  }$  &     346$^{+   123}_{-    97  }$ \\
 A3158    & 0.0590  &   $^b$5.77$^{+  0.10}_{-  0.05  }$  &   0.661$^{+ 0.025}_{- 0.022  }$  &    268$^{+    19}_{-    18  }$  &   0.39$^{+  0.02}_{-  0.01  }$  &   2.11$^{+  0.08}_{-  0.08  }$  &       0$^{+     0}_{-     0  }$ \\
 A3266    & 0.0594  &   $^a$7.70$^{+  0.80}_{-  0.80  }$  &   0.796$^{+ 0.020}_{- 0.019  }$  &    564$^{+    20}_{-    19  }$  &   0.28$^{+  0.01}_{-  0.01  }$  &   3.44$^{+  0.30}_{-  0.30  }$  &       0$^{+     0}_{-     0  }$ \\
 A3376    & 0.0455  &   $^a$4.30$^{+  0.60}_{-  0.60  }$  &   1.054$^{+ 0.101}_{- 0.083  }$  &    754$^{+    68}_{-    60  }$  &   0.10$^{+  0.01}_{-  0.00  }$  &   6.87$^{+  0.83}_{-  0.78  }$  &       0$^{+     0}_{-     0  }$ \\
 A3391    & 0.0531  &   $^a$5.70$^{+  0.70}_{-  0.70  }$  &   0.579$^{+ 0.026}_{- 0.024  }$  &    234$^{+    23}_{-    21  }$  &   0.43$^{+  0.08}_{-  0.06  }$  &   1.91$^{+  0.34}_{-  0.34  }$  &       0$^{+     0}_{-     0  }$ \\
 A3395s   & 0.0498  &   $^a$4.80$^{+  0.40}_{-  0.40  }$  &   0.964$^{+ 0.275}_{- 0.167  }$  &    604$^{+   172}_{-   117  }$  &   0.13$^{+  0.02}_{-  0.02  }$  &   5.82$^{+  1.01}_{-  0.92  }$  &       0$^{+     0}_{-     0  }$ \\
 A3526    & 0.0103  &   3.68$^{+  0.06}_{-  0.06  }$  &   0.495$^{+ 0.011}_{- 0.010  }$  &     37$^{+     4}_{-     4  }$  &   1.83$^{+  0.13}_{-  0.12  }$  &   0.36$^{+  0.03}_{-  0.02  }$  &      24$^{+     6}_{-     5  }$ \\
 A3558    & 0.0480  &   $^a$5.50$^{+  0.30}_{-  0.30  }$  &   0.580$^{+ 0.006}_{- 0.005  }$  &    223$^{+     5}_{-     5  }$  &   0.46$^{+  0.01}_{-  0.01  }$  &   1.77$^{+  0.09}_{-  0.09  }$  &       0$^{+     0}_{-     0  }$ \\
 A3562    & 0.0499  &   $^b$5.16$^{+  0.16}_{-  0.16  }$  &   0.472$^{+ 0.006}_{- 0.006  }$  &     98$^{+     5}_{-     5  }$  &   0.58$^{+  0.02}_{-  0.02  }$  &   1.33$^{+  0.07}_{-  0.06  }$  &       0$^{+     0}_{-     0  }$ \\
 A3571    & 0.0397  &   $^a$6.90$^{+  0.30}_{-  0.30  }$  &   0.613$^{+ 0.010}_{- 0.010  }$  &    181$^{+     6}_{-     6  }$  &   1.09$^{+  0.09}_{-  0.08  }$  &   0.84$^{+  0.08}_{-  0.07  }$  &      35$^{+    11}_{-    10  }$ \\
 A3581    & 0.0214  &   1.83$^{+  0.04}_{-  0.04  }$  &   0.543$^{+ 0.024}_{- 0.022  }$  &     35$^{+     4}_{-     4  }$  &   1.61$^{+  0.12}_{-  0.10  }$  &   0.25$^{+  0.02}_{-  0.02  }$  &      49$^{+    12}_{-    11  }$ \\
 A3667    & 0.0560  &   $^a$7.00$^{+  0.60}_{-  0.60  }$  &   0.541$^{+ 0.008}_{- 0.008  }$  &    279$^{+    10}_{-     9  }$  &   0.33$^{+  0.01}_{-  0.01  }$  &   2.81$^{+  0.21}_{-  0.17  }$  &       0$^{+     0}_{-     0  }$ \\
 A4038    & 0.0283  &   $^b$3.15$^{+  0.03}_{-  0.03  }$  &   0.541$^{+ 0.009}_{- 0.008  }$  &     58$^{+     3}_{-     3  }$  &   1.49$^{+  0.09}_{-  0.09  }$  &   0.40$^{+  0.02}_{-  0.02  }$  &      68$^{+    14}_{-    12  }$ \\
 A4059    & 0.0460  &   $^a$4.10$^{+  0.30}_{-  0.30  }$  &   0.582$^{+ 0.010}_{- 0.010  }$  &     89$^{+     5}_{-     5  }$  &   1.18$^{+  0.08}_{-  0.08  }$  &   0.58$^{+  0.05}_{-  0.05  }$  &      69$^{+    20}_{-    15  }$ \\
 COMA     & 0.0232  &   $^d$8.38$^{+  0.34}_{-  0.34  }$  &   0.654$^{+ 0.019}_{- 0.021  }$  &    343$^{+    22}_{-    20  }$  &   0.30$^{+  0.06}_{-  0.06  }$  &   3.45$^{+  0.85}_{-  0.56  }$  &       0$^{+     0}_{-     0  }$ \\
 EXO0422  & 0.0390  &   $^c$2.90$^{+  0.90}_{-  0.60  }$  &   0.722$^{+ 0.104}_{- 0.071  }$  &    142$^{+    40}_{-    30  }$  &   0.66$^{+  0.11}_{-  0.08  }$  &   0.85$^{+  0.23}_{-  0.19  }$  &      48$^{+    59}_{-    38  }$ \\
 FORNAX   & 0.0046  &   $^d$1.20$^{+  0.04}_{-  0.04  }$  &   0.804$^{+ 0.098}_{- 0.084  }$  &    173$^{+    17}_{-    15  }$  &   0.09$^{+  0.01}_{-  0.01  }$  &   2.50$^{+  0.20}_{-  0.18  }$  &       0$^{+     0}_{-     0  }$ \\
 HYDRA-A  & 0.0538  &   $^a$3.80$^{+  0.20}_{-  0.20  }$  &   0.573$^{+ 0.003}_{- 0.003  }$  &     50$^{+     1}_{-     1  }$  &   3.58$^{+  0.37}_{-  0.34  }$  &   0.18$^{+  0.02}_{-  0.02  }$  &     293$^{+   150}_{-    84  }$ \\
 IIIZw54  & 0.0311  &  (2.16$^{+  0.35}_{-  0.30  }$) &   0.887$^{+ 0.320}_{- 0.151  }$  &    289$^{+   123}_{-    72  }$  &   0.20$^{+  0.05}_{-  0.03  }$  &   2.31$^{+  0.55}_{-  0.56  }$  &       0$^{+     0}_{-     0  }$ \\
 MKW3S    & 0.0450  &   $^a$3.50$^{+  0.20}_{-  0.20  }$  &   0.581$^{+ 0.008}_{- 0.007  }$  &     66$^{+     2}_{-     2  }$  &   1.89$^{+  0.25}_{-  0.25  }$  &   0.33$^{+  0.06}_{-  0.04  }$  &     121$^{+    76}_{-    44  }$ \\
 MKW4     & 0.0200  &   $^d$1.71$^{+  0.09}_{-  0.09  }$  &   0.440$^{+ 0.004}_{- 0.005  }$  &     11$^{+     0}_{-     0  }$  &   2.92$^{+  0.11}_{-  0.09  }$  &   0.13$^{+  0.01}_{-  0.01  }$  &      16$^{+     2}_{-     2  }$ \\
 MKW8     & 0.0270  &   3.29$^{+  0.23}_{-  0.22  }$  &   0.511$^{+ 0.098}_{- 0.059  }$  &    106$^{+    70}_{-    42  }$  &   0.26$^{+  0.11}_{-  0.05  }$  &   2.31$^{+  0.62}_{-  0.69  }$  &       0$^{+     0}_{-     0  }$ \\
 NGC1550  & 0.0123  &   1.43$^{+  0.04}_{-  0.03  }$  &   0.554$^{+ 0.049}_{- 0.037  }$  &     45$^{+    15}_{-    10  }$  &   0.75$^{+  0.14}_{-  0.09  }$  &   0.41$^{+  0.06}_{-  0.06  }$  &      20$^{+    10}_{-     8  }$ \\
 NGC4636  & 0.0037  &   0.76$^{+  0.01}_{-  0.01  }$  &   0.491$^{+ 0.032}_{- 0.027  }$  &      6$^{+     2}_{-     1  }$  &   1.68$^{+  0.40}_{-  0.24  }$  &   0.07$^{+  0.01}_{-  0.01  }$  &       2$^{+     2}_{-     1  }$ \\
 NGC5044  & 0.0090  &   $^d$1.07$^{+  0.01}_{-  0.01  }$  &   0.524$^{+ 0.002}_{- 0.003  }$  &     11$^{+     0}_{-     0  }$  &   3.45$^{+  0.03}_{-  0.03  }$  &   0.05$^{+  0.00}_{-  0.00  }$  &      28$^{+     1}_{-     1  }$ \\
 NGC507   & 0.0165  &   $^d$1.26$^{+  0.07}_{-  0.07  }$  &   0.444$^{+ 0.005}_{- 0.005  }$  &     19$^{+     1}_{-     1  }$  &   1.16$^{+  0.04}_{-  0.04  }$  &   0.22$^{+  0.01}_{-  0.01  }$  &      14$^{+     2}_{-     2  }$ \\
 S1101    & 0.0580  &   $^c$3.00$^{+  1.20}_{-  0.70  }$  &   0.639$^{+ 0.006}_{- 0.007  }$  &     55$^{+     1}_{-     1  }$  &   2.90$^{+  0.19}_{-  0.18  }$  &   0.20$^{+  0.06}_{-  0.04  }$  &     299$^{+   179}_{-   112  }$ \\
 ZwCl1215 & 0.0750  &  (5.58$^{+  0.89}_{-  0.78  }$) &   0.819$^{+ 0.038}_{- 0.034  }$  &    431$^{+    27}_{-    25  }$  &   0.27$^{+  0.01}_{-  0.01  }$  &   3.00$^{+  0.36}_{-  0.33  }$  &       0$^{+     0}_{-     0  }$ \\
\hline\noalign{\tiny{Note: $^{a)}$: Markevitch 1998. 
$^{b)}$: White 2000. $^{c)}$: Edge \& Stewart 1991. $T_m$ with a, b or c has no
cooling flow correction. $^{d)}$: Fukazawa et al. 1998 with cooling flow correction. 
$T_m$ in a bracket is estimated from the $L_X$--$T$ relation given by Markevitch
(1998). Others are from Reiprich \& B\"ohringer (2002) and references therein.}}
\end{tabular}
\end{table*}

\begin{table*}
\renewcommand{\arraystretch}{0.9}
\caption[]{Cluster properties of the extended cluster sample.}
\begin{tabular}{cccccccc}
\hline
\hline\noalign{}
 Name  & z   & $T_m$ (keV) & $\beta$ & $r_c$ (kpc) & $n_{center}$ (10$^{-2}$ cm$^{-3}$) & $t_{cool}$ ($10^{10}$ yr) &
 $\dot{M}$ (M$_{\sun}/yr$)    \\
\hline\noalign{}
 3C129    & 0.0223  &   $^c$5.60$^{+  0.70}_{-  0.60  }$  &   0.601$^{+ 0.260}_{- 0.131  }$  &    318$^{+   178}_{-   107  }$  &   0.18$^{+  0.07}_{-  0.04  }$  &   4.71$^{+  1.62}_{-  1.39  }$  &       0$^{+     0}_{-     0  }$ \\
 A0539    & 0.0288  &   $^d$3.24$^{+  0.09}_{-  0.09  }$  &   0.561$^{+ 0.020}_{- 0.018  }$  &    147$^{+    13}_{-    12  }$  &   0.72$^{+  0.12}_{-  0.09  }$  &   0.84$^{+  0.13}_{-  0.12  }$  &       3$^{+     1}_{-     1  }$ \\
 A0548e   & 0.0410  &   $^b$3.10$^{+  0.10}_{-  0.10  }$  &   0.480$^{+ 0.013}_{- 0.013  }$  &    118$^{+    12}_{-    11  }$  &   0.28$^{+  0.02}_{-  0.01  }$  &   2.09$^{+  0.11}_{-  0.12  }$  &       0$^{+     0}_{-     0  }$ \\
 A0548w   & 0.0424  &  (1.20$^{+  0.19}_{-  0.17  }$) &   0.666$^{+ 0.194}_{- 0.111  }$  &    198$^{+    89}_{-    61  }$  &   0.10$^{+  0.03}_{-  0.02  }$  &   2.28$^{+  0.67}_{-  0.59  }$  &       0$^{+     0}_{-     0  }$ \\
 A0644    & 0.0704  &   $^a$7.10$^{+  0.60}_{-  0.60  }$  &   0.700$^{+ 0.011}_{- 0.011  }$  &    202$^{+     6}_{-     6  }$  &   0.78$^{+  0.01}_{-  0.01  }$  &   1.18$^{+  0.08}_{-  0.08  }$  &      16$^{+    29}_{-    15  }$ \\
 A1413    & 0.1427  &   $^b$7.32$^{+  0.26}_{-  0.24  }$  &   0.660$^{+ 0.017}_{- 0.015  }$  &    178$^{+    12}_{-    11  }$  &   1.24$^{+  0.07}_{-  0.07  }$  &   0.72$^{+  0.05}_{-  0.04  }$  &     190$^{+    40}_{-    32  }$ \\
 A1689    & 0.1840  &   $^b$9.23$^{+  0.28}_{-  0.28  }$  &   0.690$^{+ 0.011}_{- 0.011  }$  &    162$^{+     6}_{-     6  }$  &   2.12$^{+  0.18}_{-  0.18  }$  &   0.47$^{+  0.04}_{-  0.04  }$  &     683$^{+   239}_{-   182  }$ \\
 A1775    & 0.0757  &   $^b$3.69$^{+  0.20}_{-  0.11  }$  &   0.673$^{+ 0.026}_{- 0.023  }$  &    259$^{+    19}_{-    17  }$  &   0.30$^{+  0.05}_{-  0.04  }$  &   2.16$^{+  0.36}_{-  0.28  }$  &       0$^{+     0}_{-     0  }$ \\
 A1800    & 0.0748  &  (4.02$^{+  0.64}_{-  0.56  }$) &   0.766$^{+ 0.308}_{- 0.139  }$  &    391$^{+   223}_{-   131  }$  &   0.18$^{+  0.07}_{-  0.04  }$  &   3.65$^{+  1.14}_{-  1.07  }$  &       0$^{+     0}_{-     0  }$ \\
 A1914    & 0.1712  &  $^b$10.53$^{+  0.51}_{-  0.50  }$  &   0.751$^{+ 0.018}_{- 0.017  }$  &    230$^{+    10}_{-    10  }$  &   1.12$^{+  0.03}_{-  0.03  }$  &   0.97$^{+  0.04}_{-  0.04  }$  &     180$^{+    50}_{-    40  }$ \\
 A2151w   & 0.0369  &   $^b$2.40$^{+  0.06}_{-  0.06  }$  &   0.564$^{+ 0.014}_{- 0.013  }$  &     68$^{+     5}_{-     4  }$  &   0.82$^{+  0.04}_{-  0.03  }$  &   0.60$^{+  0.03}_{-  0.03  }$  &      30$^{+     3}_{-     3  }$ \\
 A2319    & 0.0564  &   $^a$9.20$^{+  0.70}_{-  0.70  }$  &   0.591$^{+ 0.013}_{- 0.012  }$  &    284$^{+    14}_{-    13  }$  &   0.51$^{+  0.05}_{-  0.05  }$  &   2.10$^{+  0.24}_{-  0.21  }$  &       0$^{+     0}_{-     0  }$ \\
 A2734    & 0.0620  &  (3.85$^{+  0.62}_{-  0.54  }$) &   0.624$^{+ 0.034}_{- 0.029  }$  &    211$^{+    26}_{-    23  }$  &   0.32$^{+  0.02}_{-  0.02  }$  &   2.04$^{+  0.28}_{-  0.27  }$  &       0$^{+     0}_{-     0  }$ \\
 A2877    & 0.0241  &   3.50$^{+  2.20}_{-  1.10  }$  &   0.566$^{+ 0.029}_{- 0.025  }$  &    189$^{+    18}_{-    16  }$  &   0.19$^{+  0.02}_{-  0.02  }$  &   3.33$^{+  1.60}_{-  0.95  }$  &       0$^{+     0}_{-     0  }$ \\
 A3395n   & 0.0498  &   $^a$4.80$^{+  0.40}_{-  0.40  }$  &   0.981$^{+ 0.619}_{- 0.244  }$  &    672$^{+   383}_{-   203  }$  &   0.10$^{+  0.04}_{-  0.02  }$  &   7.33$^{+  2.40}_{-  2.03  }$  &       0$^{+     0}_{-     0  }$ \\
 A3528n   & 0.0540  &   3.40$^{+  1.66}_{-  0.64  }$  &   0.621$^{+ 0.034}_{- 0.030  }$  &    177$^{+    16}_{-    15  }$  &   0.34$^{+  0.02}_{-  0.02  }$  &   1.81$^{+  0.61}_{-  0.28  }$  &       0$^{+     0}_{-     0  }$ \\
 A3528s   & 0.0551  &   3.15$^{+  0.89}_{-  0.59  }$  &   0.463$^{+ 0.013}_{- 0.012  }$  &    100$^{+     8}_{-     8  }$  &   0.48$^{+  0.03}_{-  0.02  }$  &   1.21$^{+  0.26}_{-  0.19  }$  &       1$^{+    11}_{-     1  }$ \\
 A3530    & 0.0544  &   3.89$^{+  0.27}_{-  0.25  }$  &   0.773$^{+ 0.114}_{- 0.085  }$  &    420$^{+    74}_{-    61  }$  &   0.13$^{+  0.01}_{-  0.01  }$  &   5.27$^{+  0.61}_{-  0.56  }$  &       0$^{+     0}_{-     0  }$ \\
 A3532    & 0.0539  &   4.58$^{+  0.19}_{-  0.17  }$  &   0.653$^{+ 0.034}_{- 0.029  }$  &    281$^{+    26}_{-    24  }$  &   0.30$^{+  0.06}_{-  0.05  }$  &   2.42$^{+  0.50}_{-  0.38  }$  &       0$^{+     0}_{-     0  }$ \\
 A3560    & 0.0495  &  (3.16$^{+  0.51}_{-  0.44  }$) &   0.566$^{+ 0.033}_{- 0.029  }$  &    255$^{+    30}_{-    27  }$  &   0.17$^{+  0.01}_{-  0.01  }$  &   3.47$^{+  0.51}_{-  0.46  }$  &       0$^{+     0}_{-     0  }$ \\
 A3627    & 0.0163  &   $^b$6.02$^{+  0.08}_{-  0.08  }$  &   0.555$^{+ 0.056}_{- 0.044  }$  &    299$^{+    56}_{-    49  }$  &   0.19$^{+  0.02}_{-  0.02  }$  &   4.51$^{+  0.44}_{-  0.45  }$  &       0$^{+     0}_{-     0  }$ \\
 A3695    & 0.0890  &  (5.29$^{+  0.85}_{-  0.74  }$) &   0.642$^{+ 0.259}_{- 0.117  }$  &    398$^{+   253}_{-   149  }$  &   0.20$^{+  0.08}_{-  0.05  }$  &   3.82$^{+  1.27}_{-  1.12  }$  &       0$^{+     0}_{-     0  }$ \\
 A3822    & 0.0760  &  (4.90$^{+  0.78}_{-  0.69  }$) &   0.639$^{+ 0.150}_{- 0.093  }$  &    350$^{+   159}_{-   111  }$  &   0.21$^{+  0.05}_{-  0.03  }$  &   3.60$^{+  0.83}_{-  0.86  }$  &       0$^{+     0}_{-     0  }$ \\
 A3827    & 0.0980  &  (7.08$^{+  1.13}_{-  0.99  }$) &   0.989$^{+ 0.410}_{- 0.192  }$  &    593$^{+   247}_{-   148  }$  &   0.26$^{+  0.06}_{-  0.05  }$  &   3.43$^{+  0.86}_{-  0.72  }$  &       0$^{+     0}_{-     0  }$ \\
 A3888    & 0.1510  &  (8.84$^{+  1.41}_{-  1.24  }$) &   0.928$^{+ 0.084}_{- 0.066  }$  &    400$^{+    45}_{-    39  }$  &   0.52$^{+  0.04}_{-  0.03  }$  &   1.92$^{+  0.26}_{-  0.25  }$  &       0$^{+     0}_{-     0  }$ \\
 A3921    & 0.0936  &   $^b$5.73$^{+  0.24}_{-  0.23  }$  &   0.762$^{+ 0.036}_{- 0.030  }$  &    328$^{+    25}_{-    23  }$  &   0.34$^{+  0.01}_{-  0.01  }$  &   2.34$^{+  0.12}_{-  0.12  }$  &       0$^{+     0}_{-     0  }$ \\
 AWM7     & 0.0172  &   $^d$3.75$^{+  0.09}_{-  0.09  }$  &   0.671$^{+ 0.027}_{- 0.025  }$  &    173$^{+    17}_{-    15  }$  &   0.60$^{+  0.05}_{-  0.05  }$  &   1.10$^{+  0.11}_{-  0.09  }$  &       6$^{+     3}_{-     4  }$ \\
 HCG94    & 0.0417  &   3.45$^{+  0.30}_{-  0.30  }$  &   0.514$^{+ 0.007}_{- 0.006  }$  &     86$^{+     4}_{-     3  }$  &   0.70$^{+  0.02}_{-  0.02  }$  &   0.89$^{+  0.07}_{-  0.07  }$  &       6$^{+     3}_{-     2  }$ \\
 IIZw108  & 0.0494  &  (3.44$^{+  0.55}_{-  0.48  }$) &   0.662$^{+ 0.167}_{- 0.097  }$  &    365$^{+   159}_{-   105  }$  &   0.14$^{+  0.04}_{-  0.02  }$  &   4.50$^{+  1.06}_{-  1.04  }$  &       0$^{+     0}_{-     0  }$ \\
 M49      & 0.0044  &   0.95$^{+  0.02}_{-  0.01  }$  &   0.592$^{+ 0.007}_{- 0.007  }$  &     10$^{+     0}_{-     0  }$  &   1.33$^{+  0.02}_{-  0.02  }$  &   0.12$^{+  0.00}_{-  0.00  }$  &       2$^{+     0}_{-     0  }$ \\
 NGC499     & 0.0147  &   0.72$^{+  0.03}_{-  0.02  }$  &   0.722$^{+ 0.034}_{- 0.030  }$  &     23$^{+     2}_{-     1  }$  &   0.94$^{+  0.10}_{-  0.09  }$  &   0.12$^{+  0.01}_{-  0.01  }$  &      11$^{+     4}_{-     3  }$ \\
 NGC5813    & 0.0064  &  (0.52$^{+  0.08}_{-  0.07  }$) &   0.766$^{+ 0.179}_{- 0.103  }$  &     24$^{+     8}_{-     5  }$  &   0.90$^{+  0.18}_{-  0.13  }$  &   0.10$^{+  0.02}_{-  0.02  }$  &       9$^{+    11}_{-     5  }$ \\
 NGC5846    & 0.0061  &   0.82$^{+  0.01}_{-  0.01  }$  &   0.599$^{+ 0.016}_{- 0.015  }$  &      6$^{+     0}_{-     0  }$  &   3.56$^{+  0.22}_{-  0.21  }$  &   0.04$^{+  0.00}_{-  0.00  }$  &       2$^{+     0}_{-     0  }$ \\
 OPHIUCHU & 0.0280  &  10.26$^{+  0.32}_{-  0.32  }$  &   0.747$^{+ 0.035}_{- 0.032  }$  &    278$^{+    23}_{-    21  }$  &   0.68$^{+  0.04}_{-  0.04  }$  &   1.72$^{+  0.12}_{-  0.10  }$  &       0$^{+     0}_{-     0  }$ \\
 PERSEUS  & 0.0183  &   $^d$6.79$^{+  0.12}_{-  0.12  }$  &   0.540$^{+ 0.006}_{- 0.004  }$  &     63$^{+     2}_{-     1  }$  &   3.25$^{+  0.06}_{-  0.05  }$  &   0.28$^{+  0.01}_{-  0.01  }$  &     481$^{+    31}_{-    32  }$ \\
 PKS0745  & 0.1028  &   $^b$7.21$^{+  0.11}_{-  0.11  }$  &   0.608$^{+ 0.006}_{- 0.006  }$  &     71$^{+     2}_{-     2  }$  &   5.70$^{+  0.14}_{-  0.15  }$  &   0.16$^{+  0.00}_{-  0.00  }$  &    1424$^{+   150}_{-   133  }$ \\
 RXJ2344  & 0.0786  &  (4.73$^{+  0.76}_{-  0.66  }$) &   0.807$^{+ 0.033}_{- 0.030  }$  &    300$^{+    19}_{-    18  }$  &   0.51$^{+  0.06}_{-  0.05  }$  &   1.43$^{+  0.24}_{-  0.20  }$  &       0$^{+     1}_{-     0  }$ \\
 S405     & 0.0613  &  (4.21$^{+  0.67}_{-  0.59  }$) &   0.664$^{+ 0.263}_{- 0.133  }$  &    458$^{+   261}_{-   158  }$  &   0.12$^{+  0.05}_{-  0.03  }$  &   5.67$^{+  1.85}_{-  1.74  }$  &       0$^{+     0}_{-     0  }$ \\
 S540     & 0.0358  &  (2.40$^{+  0.38}_{-  0.34  }$) &   0.641$^{+ 0.073}_{- 0.051  }$  &    130$^{+    38}_{-    28  }$  &   0.40$^{+  0.07}_{-  0.05  }$  &   1.23$^{+  0.23}_{-  0.24  }$  &       1$^{+     7}_{-     1  }$ \\
 S636     & 0.0116  &  (1.18$^{+  0.19}_{-  0.17  }$) &   0.752$^{+ 0.217}_{- 0.123  }$  &    343$^{+   130}_{-    86  }$  &   0.07$^{+  0.01}_{-  0.01  }$  &   3.06$^{+  0.77}_{-  0.65  }$  &       0$^{+     0}_{-     0  }$ \\
 TRIANGUL & 0.0510  &   $^a$9.50$^{+  0.70}_{-  0.70  }$  &   0.610$^{+ 0.010}_{- 0.010  }$  &    278$^{+    10}_{-     9  }$  &   0.55$^{+  0.01}_{-  0.01  }$  &   1.98$^{+  0.12}_{-  0.12  }$  &       0$^{+     0}_{-     0  }$ \\
 UGC03957 & 0.0340  &  (2.58$^{+  0.41}_{-  0.36  }$) &   0.740$^{+ 0.133}_{- 0.086  }$  &    142$^{+    44}_{-    33  }$  &   0.48$^{+  0.08}_{-  0.06  }$  &   1.09$^{+  0.21}_{-  0.20  }$  &       8$^{+    11}_{-     8  }$ \\
 ZwCl1742 & 0.0757  &  (5.23$^{+  0.84}_{-  0.73  }$) &   0.717$^{+ 0.073}_{- 0.053  }$  &    231$^{+    45}_{-    38  }$  &   0.60$^{+  0.11}_{-  0.09  }$  &   1.29$^{+  0.28}_{-  0.23  }$  &       0$^{+    23}_{-     0  }$ \\
\hline\noalign{Note: The symbols in $T_m$ have the same meanings as in Table 1.}
\end{tabular}
\end{table*}

\renewcommand{\arraystretch}{0.65}
\begin{table*}
\caption[]{Fit parameters with a double $\beta$ model. }
\begin{tabular}{ccccccccc}
\hline
\hline\noalign{}
 Name  & $S_{01}$ (10$^{-6}$ cts/s/pixel$^2$) & $S_{02}$ (10$^{-6}$ cts/s/pixel$^2$)
  & $\beta_1$ &  $\beta_2$ & $r_{c1}$ (kpc) &
  $r_{c2}$ (kpc) & $\chi^2_s$ &  $\chi^2_d$ \\
\hline\noalign{}
 A0085    &  27.01$\pm{  0.99  }$  &   1.70$\pm{  0.24  }$  &   0.60$\pm{  0.03  }$  &   0.73$\pm{  0.03  }$  &     58$\pm{     3  }$  &    385$\pm{    28  }$  &   5.32   &   1.90 \\
 A0119    &   0.50$\pm{  0.03  }$  &   0.16$\pm{  0.03  }$  &   0.76$\pm{  0.11  }$  &   1.46$\pm{  0.24  }$  &    399$\pm{    45  }$  &   1511$\pm{   207  }$  &   1.75   &   1.47 \\
 A0133    &  23.96$\pm{  0.82  }$  &   0.87$\pm{  0.06  }$  &   0.65$\pm{  0.02  }$  &   0.78$\pm{  0.02  }$  &     42$\pm{     1  }$  &    321$\pm{    13  }$  &   4.14   &   1.71 \\
 A0401    &   3.93$\pm{  0.30  }$  &   0.51$\pm{  0.30  }$  &   0.69$\pm{  0.08  }$  &   0.66$\pm{  0.03  }$  &    239$\pm{    23  }$  &    525$\pm{    93  }$  &   1.19   &   1.17 \\
 A0478    &  37.46$\pm{  1.30  }$  &   4.02$\pm{  0.71  }$  &   0.68$\pm{  0.04  }$  &   0.71$\pm{  0.02  }$  &     72$\pm{     4  }$  &    253$\pm{    14  }$  &   3.53   &   1.50 \\
 A0496    &  37.20$\pm{  1.87  }$  &   1.40$\pm{  0.33  }$  &   0.59$\pm{  0.05  }$  &   0.69$\pm{  0.04  }$  &     30$\pm{     3  }$  &    257$\pm{    31  }$  &   3.97   &   1.07 \\
 A1367    &   0.30$\pm{  0.01  }$  &   0.22$\pm{  0.01  }$  &   0.96$\pm{  0.10  }$  &   1.51$\pm{  0.06  }$  &    291$\pm{    23  }$  &    976$\pm{    31  }$  &   1.51   &   1.14 \\
 A1644    &   1.16$\pm{  0.24  }$  &   0.18$\pm{  0.06  }$  &   0.83$\pm{  0.30  }$  &   2.38$\pm{  0.81  }$  &    274$\pm{    92  }$  &   2169$\pm{   532  }$  &   0.97   &   0.89 \\
 A1651    &   7.39$\pm{  0.47  }$  &   2.17$\pm{  0.24  }$  &   0.75$\pm{  0.07  }$  &   0.76$\pm{  0.02  }$  &    120$\pm{    11  }$  &    356$\pm{    17  }$  &   1.46   &   1.10 \\
 A1795    &  41.44$\pm{  0.60  }$  &   1.75$\pm{  0.16  }$  &   0.72$\pm{  0.02  }$  &   0.89$\pm{  0.02  }$  &     79$\pm{     2  }$  &    432$\pm{    19  }$  &  10.04   &   1.99 \\
 A2029    &  55.67$\pm{  1.98  }$  &   5.81$\pm{  1.16  }$  &   0.63$\pm{  0.03  }$  &   0.65$\pm{  0.02  }$  &     62$\pm{     3  }$  &    213$\pm{    15  }$  &   2.61   &   1.40 \\
 A2052    &  23.26$\pm{  1.43  }$  &   3.78$\pm{  0.46  }$  &   2.10$\pm{  1.02  }$  &   0.66$\pm{  0.02  }$  &     84$\pm{    25  }$  &    140$\pm{    12  }$  &   2.70   &   1.45 \\
 A2063    &   4.71$\pm{  0.00  }$  &   0.51$\pm{  0.00  }$  &   0.49$\pm{  0.00  }$  &   2.02$\pm{  0.00  }$  &     55$\pm{     0  }$  &    640$\pm{     0  }$  &   1.97   &   1.17 \\
 A2142    &  19.59$\pm{  0.69  }$  &   0.86$\pm{  0.19  }$  &   0.67$\pm{  0.04  }$  &   1.01$\pm{  0.13  }$  &    140$\pm{     8  }$  &    893$\pm{   130  }$  &   2.55   &   1.39 \\
 A2255    &   0.35$\pm{  0.05  }$  &   0.56$\pm{  0.05  }$  &   1.15$\pm{  0.32  }$  &   0.90$\pm{  0.03  }$  &    469$\pm{    93  }$  &    778$\pm{    31  }$  &   1.38   &   1.36 \\
 A2589    &   3.45$\pm{  0.28  }$  &   0.51$\pm{  0.23  }$  &   0.66$\pm{  0.10  }$  &   0.74$\pm{  0.07  }$  &     95$\pm{    13  }$  &    311$\pm{    56  }$  &   1.38   &   1.23 \\
 A2634    &   0.42$\pm{  0.04  }$  &   0.13$\pm{  0.01  }$  &   0.47$\pm{  0.03  }$  &   1.89$\pm{  0.18  }$  &     80$\pm{    10  }$  &   1189$\pm{    77  }$  &   1.77   &   1.25 \\
 A2657    &   2.34$\pm{  0.00  }$  &   0.25$\pm{  0.00  }$  &   0.89$\pm{  0.00  }$  &   1.27$\pm{  0.00  }$  &    148$\pm{     0  }$  &    796$\pm{     0  }$  &   3.05   &   1.68 \\
 A3112    &  41.34$\pm{  2.59  }$  &   3.11$\pm{  1.41  }$  &   0.63$\pm{  0.06  }$  &   0.62$\pm{  0.03  }$  &     51$\pm{     5  }$  &    164$\pm{    24  }$  &   1.62   &   1.44 \\
 A3266    &   1.24$\pm{  0.05  }$  &   0.76$\pm{  0.04  }$  &   1.20$\pm{  0.11  }$  &   1.27$\pm{  0.03  }$  &    450$\pm{    29  }$  &   1162$\pm{    28  }$  &   2.62   &   1.76 \\
 A3391    &   0.87$\pm{  0.16  }$  &   0.66$\pm{  0.08  }$  &   0.50$\pm{  0.06  }$  &   0.66$\pm{  0.03  }$  &     71$\pm{    19  }$  &    335$\pm{    24  }$  &   1.30   &   1.26 \\
 A3526    &   8.97$\pm{  0.79  }$  &   0.34$\pm{  0.07  }$  &   0.57$\pm{  0.03  }$  &   0.70$\pm{  0.05  }$  &     33$\pm{     3  }$  &    272$\pm{    32  }$  &   1.22   &   1.11 \\
 A3558    &   3.13$\pm{  0.05  }$  &   0.20$\pm{  0.04  }$  &   0.68$\pm{  0.03  }$  &   1.17$\pm{  0.11  }$  &    232$\pm{     8  }$  &   1198$\pm{   119  }$  &   3.66   &   2.90 \\
 A3562    &   2.84$\pm{  0.10  }$  &   0.07$\pm{  0.03  }$  &   0.52$\pm{  0.02  }$  &   1.26$\pm{  0.27  }$  &    103$\pm{     6  }$  &   1341$\pm{   273  }$  &   2.10   &   1.78 \\
 A3571    &   5.00$\pm{  0.40  }$  &   4.39$\pm{  0.22  }$  &   0.82$\pm{  0.13  }$  &   0.68$\pm{  0.01  }$  &     94$\pm{    13  }$  &    256$\pm{     6  }$  &   1.72   &   1.13 \\
 A3667    &   2.10$\pm{  0.04  }$  &   0.20$\pm{  0.01  }$  &   0.89$\pm{  0.03  }$  &   1.70$\pm{  0.09  }$  &    402$\pm{    12  }$  &   2375$\pm{    95  }$  &   2.53   &   1.63 \\
 A4038    &  10.37$\pm{  0.58  }$  &   0.49$\pm{  0.29  }$  &   0.58$\pm{  0.04  }$  &   0.70$\pm{  0.12  }$  &     53$\pm{     4  }$  &    241$\pm{    69  }$  &   1.27   &   1.18 \\
 A4059    &   8.67$\pm{  0.46  }$  &   0.43$\pm{  0.18  }$  &   0.64$\pm{  0.06  }$  &   0.90$\pm{  0.19  }$  &     82$\pm{     7  }$  &    438$\pm{   112  }$  &   1.32   &   1.06 \\
 COMA     &   1.13$\pm{  0.54  }$  &   1.35$\pm{  0.52  }$  &   0.57$\pm{  0.05  }$  &   0.90$\pm{  0.23  }$  &    329$\pm{    66  }$  &    444$\pm{   104  }$  &   1.09   &   1.09 \\
 HYDRA-A  &  40.63$\pm{  1.40  }$  &   5.67$\pm{  0.43  }$  &   1.84$\pm{  0.42  }$  &   0.73$\pm{  0.01  }$  &     98$\pm{    14  }$  &    183$\pm{     8  }$  &   6.24   &   1.81 \\
 MKW3S    &  12.40$\pm{  0.92  }$  &   3.84$\pm{  0.62  }$  &   1.42$\pm{  0.41  }$  &   0.68$\pm{  0.02  }$  &     91$\pm{    17  }$  &    152$\pm{    13  }$  &   2.07   &   1.54 \\
 S1101    &  34.57$\pm{  1.35  }$  &   0.58$\pm{  0.30  }$  &   0.79$\pm{  0.08  }$  &   0.96$\pm{  0.16  }$  &     66$\pm{     6  }$  &    381$\pm{   104  }$  &   1.67   &   1.33 \\
 A0539    &   1.53$\pm{  0.19  }$  &   0.41$\pm{  0.08  }$  &   0.53$\pm{  0.09  }$  &   0.75$\pm{  0.08  }$  &     42$\pm{     9  }$  &    313$\pm{    40  }$  &   1.77   &   1.19 \\
 A1413    &  10.29$\pm{  0.59  }$  &   1.11$\pm{  0.20  }$  &   0.80$\pm{  0.07  }$  &   0.91$\pm{  0.05  }$  &    155$\pm{    12  }$  &    559$\pm{    47  }$  &   1.73   &   1.45 \\
 A1689    &  23.46$\pm{  1.39  }$  &   2.53$\pm{  0.82  }$  &   0.88$\pm{  0.14  }$  &   0.91$\pm{  0.05  }$  &    152$\pm{    19  }$  &    471$\pm{    56  }$  &   1.73   &   1.24 \\
 A1775    &   1.48$\pm{  0.08  }$  &   0.18$\pm{  0.04  }$  &   2.05$\pm{  0.69  }$  &   1.70$\pm{  0.53  }$  &    540$\pm{   114  }$  &   1443$\pm{   382  }$  &   1.62   &   1.46 \\
 A2319    &   3.59$\pm{  0.23  }$  &   0.76$\pm{  0.19  }$  &   1.06$\pm{  0.19  }$  &   0.82$\pm{  0.06  }$  &    383$\pm{    48  }$  &    874$\pm{   120  }$  &   1.34   &   1.24 \\
 A2877    &   0.28$\pm{  0.02  }$  &   0.15$\pm{  0.01  }$  &   3.58$\pm{  0.89  }$  &   1.23$\pm{  0.07  }$  &    323$\pm{    47  }$  &    606$\pm{    28  }$  &   2.79   &   2.04 \\
 A3532    &   0.91$\pm{  0.11  }$  &   0.27$\pm{  0.09  }$  &   0.74$\pm{  0.24  }$  &   1.09$\pm{  0.21  }$  &    193$\pm{    48  }$  &    761$\pm{   161  }$  &   1.31   &   1.12 \\
 A3888    &   3.90$\pm{  0.22  }$  &   0.30$\pm{  0.15  }$  &   1.39$\pm{  0.19  }$  &   1.71$\pm{  0.39  }$  &    503$\pm{    47  }$  &   1341$\pm{   271  }$  &   0.90   &   0.90 \\
 AWM7     &   2.53$\pm{  0.22  }$  &   0.56$\pm{  0.11  }$  &   0.78$\pm{  0.11  }$  &   0.88$\pm{  0.05  }$  &    125$\pm{    16  }$  &    406$\pm{    39  }$  &   1.40   &   1.33 \\
 HCG94    &   2.26$\pm{  0.09  }$  &   0.41$\pm{  0.04  }$  &   0.53$\pm{  0.01  }$  &   0.58$\pm{  0.01  }$  &     59$\pm{     3  }$  &    199$\pm{    10  }$  &   2.43   &   2.20 \\
 NGC507   &   1.20$\pm{  0.30  }$  &   3.37$\pm{  0.34  }$  &   0.76$\pm{  0.04  }$  &   4.29$\pm{  1.21  }$  &     41$\pm{     4  }$  &     73$\pm{    11  }$  &   2.41   &   2.40 \\
 NGC5846  &  10.88$\pm{  0.97  }$  &   1.48$\pm{  0.18  }$  &   0.51$\pm{  0.01  }$  &   4.78$\pm{  1.07  }$  &      3$\pm{     0  }$  &     55$\pm{     7  }$  &   1.78   &   1.53 \\
 OPHIUCHU &   4.86$\pm{  0.24  }$  &   0.35$\pm{  0.10  }$  &   1.04$\pm{  0.11  }$  &   1.40$\pm{  0.22  }$  &    328$\pm{    26  }$  &   1190$\pm{   177  }$  &   1.41   &   1.37 \\
 PKS0745  &  53.06$\pm{  1.73  }$  &   2.02$\pm{  0.39  }$  &   0.70$\pm{  0.02  }$  &   0.65$\pm{  0.01  }$  &     72$\pm{     2  }$  &    235$\pm{    16  }$  &   1.17   &   1.06 \\
 RXJ2344  &   1.41$\pm{  0.20  }$  &   1.41$\pm{  0.14  }$  &   0.72$\pm{  0.13  }$  &   0.92$\pm{  0.04  }$  &    128$\pm{    27  }$  &    400$\pm{    20  }$  &   1.76   &   1.70 \\
 TRIANGUL &   3.36$\pm{  0.09  }$  &   0.66$\pm{  0.06  }$  &   0.71$\pm{  0.03  }$  &   0.80$\pm{  0.02  }$  &    248$\pm{     9  }$  &    700$\pm{    30  }$  &   1.57   &   1.28 \\
 ZwCl1742 &   3.50$\pm{  0.57  }$  &   0.58$\pm{  0.33  }$  &   0.83$\pm{  0.21  }$  &   0.93$\pm{  0.16  }$  &    194$\pm{    45  }$  &    547$\pm{   135  }$  &   1.11   &   1.11 \\
\hline\noalign{Note: $\chi^2_s$ and 
$\chi^2_d$ are the reduced $\chi^2$ values for the single $\beta$ model 
and the double $\beta$ model, respectively. $S_1$ and
$S_2$ are the surface brightness. Note $1 \arcmin=120$ pixels.}
\end{tabular}
\end{table*}

\renewcommand{\arraystretch}{0.65}
\begin{table*}
\caption[]{Cluster properties for the scaling relations.}
\vskip -0.2cm
\begin{tabular}{cccccccc}
\hline\noalign{}
 Name  & $T_h$  & $L_X$  & $M_{500}$  & $r_{500}$  & $f_{gas}$
 & $M_{gas,500}$  & $L_{nir}$ \\
    & (keV)  & ($10^{44}$ erg/s, 0.1-2.4keV) &  ($10^{14}$
 M$_{\sun}$) & (Mpc) &   &  ($10^{13}$ M$_{\sun}$)
  & ($h_{70}^{-2} \, 10^{12}$ L$_{\sun}$) \\
\hline\noalign{}
 2A0335   &   3.64$^{+  0.09}_{-  0.08  }$  &   4.64 $\pm{  0.04  }$  &   2.79$^{+  1.09}_{-  1.63  }$  &   1.20$^{+  0.14}_{-  0.30  }$  &   0.19$^{+  0.13}_{-  0.03  }$  &   5.42$^{+  0.80}_{-  1.74  }$  &   3.44 $\pm{  0.07  }$ \\
 A0085    &   6.51$^{+  0.16}_{-  0.23  }$  &   9.67 $\pm{  0.11  }$  &   8.08$^{+  1.57}_{-  3.51  }$  &   1.68$^{+  0.10}_{-  0.29  }$  &   0.21$^{+  0.09}_{-  0.02  }$  &  16.66$^{+  1.49}_{-  3.54  }$  &   5.99 $\pm{  0.06  }$ \\
 A0119    &   5.69$^{+  0.24}_{-  0.28  }$  &   3.34 $\pm{  0.05  }$  &   8.98$^{+  1.20}_{-  2.59  }$  &   1.76$^{+  0.07}_{-  0.19  }$  &   0.15$^{+  0.04}_{-  0.01  }$  &  13.24$^{+  0.75}_{-  1.79  }$  &   6.46 $\pm{  0.07  }$ \\
 A0133    &   3.97$^{+  0.28}_{-  0.27  }$  &   2.85 $\pm{  0.04  }$  &   4.30$^{+  1.00}_{-  2.08  }$  &   1.36$^{+  0.10}_{-  0.27  }$  &   0.14$^{+  0.07}_{-  0.02  }$  &   6.01$^{+  0.51}_{-  1.36  }$  &   3.72 $\pm{  0.04  }$ \\
 A0262    &   2.25$^{+  0.06}_{-  0.06  }$  &   0.98 $\pm{  0.06  }$  &   0.94$^{+  0.05}_{-  0.12  }$  &   0.85$^{+  0.02}_{-  0.04  }$  &   0.21$^{+  0.05}_{-  0.05  }$  &   1.93$^{+  0.39}_{-  0.48  }$  &   2.07 $\pm{  0.07  }$ \\
 A0399    &   6.46$^{+  0.38}_{-  0.36  }$  &   7.13 $\pm{  0.62  }$  &   7.74$^{+  2.15}_{-  2.32  }$  &   1.63$^{+  0.14}_{-  0.18  }$  &   0.19$^{+  0.11}_{-  0.07  }$  &  14.79$^{+  3.43}_{-  5.62  }$  &  - \\
 A0400    &   2.43$^{+  0.13}_{-  0.12  }$  &   0.65 $\pm{  0.01  }$  &   1.33$^{+  0.11}_{-  0.16  }$  &   0.95$^{+  0.02}_{-  0.04  }$  &   0.16$^{+  0.02}_{-  0.01  }$  &   2.11$^{+  0.13}_{-  0.20  }$  &   2.34 $\pm{  0.06  }$ \\
 A0401    &   7.19$^{+  0.28}_{-  0.24  }$  &  12.41 $\pm{  0.22  }$  &   8.38$^{+  1.22}_{-  2.84  }$  &   1.67$^{+  0.08}_{-  0.21  }$  &   0.25$^{+  0.08}_{-  0.02  }$  &  20.55$^{+  2.75}_{-  4.27  }$  &  - \\
 A0478    &   6.91$^{+  0.40}_{-  0.36  }$  &  17.44 $\pm{  0.18  }$  &   8.85$^{+  3.00}_{-  4.69  }$  &   1.68$^{+  0.17}_{-  0.37  }$  &   0.22$^{+  0.14}_{-  0.04  }$  &  19.10$^{+  2.18}_{-  4.72  }$  &   7.96 $\pm{  0.09  }$ \\
 A0496    &   4.59$^{+  0.10}_{-  0.10  }$  &   3.77 $\pm{  0.05  }$  &   4.81$^{+  0.89}_{-  2.11  }$  &   1.44$^{+  0.08}_{-  0.25  }$  &   0.16$^{+  0.08}_{-  0.02  }$  &   7.77$^{+  0.77}_{-  1.65  }$  &   3.91 $\pm{  0.05  }$ \\
 A0576    &   3.83$^{+  0.16}_{-  0.15  }$  &   1.86 $\pm{  0.21  }$  &   4.61$^{+  3.25}_{-  2.39  }$  &   1.42$^{+  0.28}_{-  0.30  }$  &   0.12$^{+  0.18}_{-  0.08  }$  &   5.43$^{+  2.11}_{-  3.53  }$  &   4.62 $\pm{  0.06  }$ \\
 A0754    &   9.00$^{+  0.35}_{-  0.34  }$  &   3.97 $\pm{  0.11  }$  &  13.86$^{+  4.40}_{-  6.43  }$  &   2.02$^{+  0.19}_{-  0.38  }$  &   0.10$^{+  0.05}_{-  0.02  }$  &  13.25$^{+  1.83}_{-  3.32  }$  &   9.28 $\pm{  0.05  }$ \\
 A1060    &   3.15$^{+  0.05}_{-  0.05  }$  &   0.56 $\pm{  0.03  }$  &   2.50$^{+  0.62}_{-  1.02  }$  &   1.19$^{+  0.09}_{-  0.19  }$  &   0.09$^{+  0.04}_{-  0.02  }$  &   2.19$^{+  0.49}_{-  0.69  }$  &   2.39 $\pm{  0.06  }$ \\
 A1367    &   3.55$^{+  0.08}_{-  0.08  }$  &   1.20 $\pm{  0.02  }$  &   7.42$^{+  1.11}_{-  2.37  }$  &   1.69$^{+  0.08}_{-  0.20  }$  &   0.09$^{+  0.03}_{-  0.01  }$  &   6.39$^{+  0.30}_{-  0.69  }$  &   3.81 $\pm{  0.05  }$ \\
 A1644    &   $^a$4.70$^{+  0.90}_{-  0.70  }$  &   3.92 $\pm{  0.34  }$  &   7.34$^{+  4.30}_{-  4.40  }$  &   1.64$^{+  0.27}_{-  0.43  }$  &   0.17$^{+  0.13}_{-  0.04  }$  &  12.53$^{+  2.71}_{-  4.28  }$  &   5.82 $\pm{  0.05  }$ \\
 A1650    &   5.68$^{+  0.30}_{-  0.27  }$  &   7.33 $\pm{  0.79  }$  &   6.53$^{+  2.17}_{-  2.43  }$  &   1.52$^{+  0.15}_{-  0.22  }$  &   0.19$^{+  0.14}_{-  0.08  }$  &  12.22$^{+  4.85}_{-  5.79  }$  &   3.09 $\pm{  0.14  }$ \\
 A1651    &   6.22$^{+  0.45}_{-  0.41  }$  &   7.85 $\pm{  0.14  }$  &   8.29$^{+  1.95}_{-  3.60  }$  &   1.65$^{+  0.12}_{-  0.28  }$  &   0.17$^{+  0.08}_{-  0.02  }$  &  14.38$^{+  1.31}_{-  2.81  }$  &   7.82 $\pm{  0.05  }$ \\
 A1736    &   3.68$^{+  0.22}_{-  0.17  }$  &   3.22 $\pm{  0.33  }$  &   2.17$^{+  0.62}_{-  0.69  }$  &   1.09$^{+  0.10}_{-  0.13  }$  &   0.25$^{+  0.09}_{-  0.10  }$  &   5.33$^{+  1.07}_{-  1.84  }$  &   5.28 $\pm{  0.10  }$ \\
 A1795    &   6.17$^{+  0.26}_{-  0.25  }$  &  10.00 $\pm{  0.07  }$  &   9.87$^{+  3.85}_{-  5.48  }$  &   1.79$^{+  0.21}_{-  0.42  }$  &   0.15$^{+  0.11}_{-  0.03  }$  &  14.51$^{+  1.50}_{-  3.46  }$  &   4.80 $\pm{  0.04  }$ \\
 A2029    &   7.93$^{+  0.39}_{-  0.36  }$  &  17.07 $\pm{  0.18  }$  &   9.95$^{+  3.29}_{-  5.16  }$  &   1.77$^{+  0.18}_{-  0.38  }$  &   0.21$^{+  0.12}_{-  0.03  }$  &  20.92$^{+  2.45}_{-  5.52  }$  &   8.12 $\pm{  0.06  }$ \\
 A2052    &   3.12$^{+  0.10}_{-  0.09  }$  &   2.37 $\pm{  0.04  }$  &   2.70$^{+  0.57}_{-  1.32  }$  &   1.19$^{+  0.08}_{-  0.24  }$  &   0.16$^{+  0.09}_{-  0.02  }$  &   4.31$^{+  1.43}_{-  0.95  }$  &   3.22 $\pm{  0.05  }$ \\
 A2063    &   3.56$^{+  0.16}_{-  0.12  }$  &   2.26 $\pm{  0.05  }$  &   2.36$^{+  0.24}_{-  0.59  }$  &   1.14$^{+  0.04}_{-  0.10  }$  &   0.19$^{+  0.03}_{-  0.01  }$  &   4.44$^{+  0.20}_{-  0.52  }$  &   3.50 $\pm{  0.04  }$ \\
 A2065    &   5.37$^{+  0.34}_{-  0.30  }$  &   5.63 $\pm{  0.55  }$  &  11.19$^{+  9.57}_{-  6.82  }$  &   1.84$^{+  0.42}_{-  0.50  }$  &   0.12$^{+  0.25}_{-  0.08  }$  &  13.68$^{+  5.07}_{-  9.46  }$  &   7.33 $\pm{  0.05  }$ \\
 A2142    &   8.46$^{+  0.53}_{-  0.49  }$  &  21.05 $\pm{  0.29  }$  &  14.33$^{+  3.64}_{-  6.83  }$  &   1.97$^{+  0.15}_{-  0.38  }$  &   0.22$^{+  0.11}_{-  0.03  }$  &  31.94$^{+  3.25}_{-  7.17  }$  &   7.20 $\pm{  0.09  }$ \\
 A2147    &   4.34$^{+  0.12}_{-  0.13  }$  &   2.87 $\pm{  0.15  }$  &   2.31$^{+  0.40}_{-  0.36  }$  &   1.13$^{+  0.06}_{-  0.06  }$  &   0.26$^{+  0.07}_{-  0.08  }$  &   5.93$^{+  1.19}_{-  1.66  }$  &   4.15 $\pm{  0.05  }$ \\
 A2163    &  10.55$^{+  1.01}_{-  0.68  }$  &  32.16 $\pm{  0.82  }$  &  16.00$^{+  3.48}_{-  4.86  }$  &   1.85$^{+  0.13}_{-  0.21  }$  &   0.27$^{+  0.08}_{-  0.03  }$  &  43.75$^{+  3.73}_{-  7.16  }$  &  - \\
 A2199    &   4.28$^{+  0.10}_{-  0.10  }$  &   4.20 $\pm{  0.12  }$  &   4.29$^{+  1.18}_{-  1.89  }$  &   1.39$^{+  0.12}_{-  0.24  }$  &   0.17$^{+  0.08}_{-  0.03  }$  &   7.35$^{+  0.91}_{-  1.73  }$  &   4.43 $\pm{  0.04  }$ \\
 A2204    &   6.38$^{+  0.23}_{-  0.23  }$  &  25.89 $\pm{  0.69  }$  &   5.82$^{+  1.91}_{-  2.98  }$  &   1.38$^{+  0.14}_{-  0.29  }$  &   0.29$^{+  0.15}_{-  0.05  }$  &  16.85$^{+  2.24}_{-  4.70  }$  &  - \\
 A2244    &   5.77$^{+  0.61}_{-  0.44  }$  &   8.30 $\pm{  0.28  }$  &   5.48$^{+  1.48}_{-  2.23  }$  &   1.42$^{+  0.12}_{-  0.23  }$  &   0.21$^{+  0.08}_{-  0.04  }$  &  11.52$^{+  1.71}_{-  2.65  }$  &  - \\
 A2255    &   5.92$^{+  0.40}_{-  0.26  }$  &   5.46 $\pm{  0.11  }$  &   7.86$^{+  0.92}_{-  1.67  }$  &   1.63$^{+  0.06}_{-  0.12  }$  &   0.19$^{+  0.03}_{-  0.01  }$  &  14.66$^{+  1.06}_{-  1.50  }$  &   8.70 $\pm{  0.08  }$ \\
 A2256    &   6.83$^{+  0.23}_{-  0.21  }$  &   9.24 $\pm{  0.22  }$  &  12.12$^{+  3.41}_{-  4.12  }$  &   1.91$^{+  0.16}_{-  0.25  }$  &   0.18$^{+  0.07}_{-  0.03  }$  &  21.78$^{+  2.19}_{-  4.09  }$  &  10.11 $\pm{  0.04  }$ \\
 A2589    &   3.38$^{+  0.13}_{-  0.13  }$  &   1.87 $\pm{  0.04  }$  &   3.24$^{+  0.54}_{-  1.40  }$  &   1.26$^{+  0.07}_{-  0.21  }$  &   0.15$^{+  0.08}_{-  0.02  }$  &   4.76$^{+  0.81}_{-  0.99  }$  &   2.51 $\pm{  0.06  }$ \\
 A2597    &   4.20$^{+  0.49}_{-  0.41  }$  &   6.75 $\pm{  0.14  }$  &   3.71$^{+  1.35}_{-  2.22  }$  &   1.26$^{+  0.14}_{-  0.33  }$  &   0.17$^{+  0.13}_{-  0.03  }$  &   6.48$^{+  0.86}_{-  2.02  }$  &  - \\
 A2634    &   3.45$^{+  0.16}_{-  0.16  }$  &   0.99 $\pm{  0.03  }$  &   4.51$^{+  0.67}_{-  1.00  }$  &   1.42$^{+  0.07}_{-  0.11  }$  &   0.11$^{+  0.02}_{-  0.01  }$  &   5.10$^{+  0.16}_{-  0.37  }$  &   4.53 $\pm{  0.08  }$ \\
 A2657    &   3.53$^{+  0.12}_{-  0.12  }$  &   1.74 $\pm{  0.03  }$  &   6.06$^{+  1.32}_{-  2.57  }$  &   1.55$^{+  0.11}_{-  0.26  }$  &   0.10$^{+  0.05}_{-  0.01  }$  &   6.26$^{+  0.39}_{-  1.08  }$  &   2.35 $\pm{  0.09  }$ \\
 A3112    &   4.72$^{+  0.37}_{-  0.25  }$  &   7.32 $\pm{  0.15  }$  &   4.36$^{+  1.26}_{-  2.25  }$  &   1.34$^{+  0.12}_{-  0.29  }$  &   0.20$^{+  0.13}_{-  0.03  }$  &   8.87$^{+  1.71}_{-  2.49  }$  &   4.11 $\pm{  0.08  }$ \\
 A3158    &   5.41$^{+  0.26}_{-  0.24  }$  &   5.61 $\pm{  0.15  }$  &   5.75$^{+  0.89}_{-  1.66  }$  &   1.49$^{+  0.07}_{-  0.16  }$  &   0.20$^{+  0.05}_{-  0.03  }$  &  11.63$^{+  1.13}_{-  2.10  }$  &   6.76 $\pm{  0.05  }$ \\
 A3266    &   7.72$^{+  0.35}_{-  0.28  }$  &   8.62 $\pm{  0.11  }$  &  19.24$^{+  4.76}_{-  7.58  }$  &   2.23$^{+  0.17}_{-  0.34  }$  &   0.14$^{+  0.05}_{-  0.02  }$  &  27.35$^{+  1.97}_{-  4.39  }$  &   7.79 $\pm{  0.06  }$ \\
 A3376    &   4.43$^{+  0.39}_{-  0.38  }$  &   2.16 $\pm{  0.05  }$  &   6.77$^{+  1.55}_{-  1.99  }$  &   1.60$^{+  0.11}_{-  0.17  }$  &   0.13$^{+  0.04}_{-  0.02  }$  &   8.70$^{+  0.86}_{-  1.50  }$  &   3.81 $\pm{  0.04  }$ \\
 A3391    &   5.89$^{+  0.45}_{-  0.33  }$  &   2.64 $\pm{  0.08  }$  &   6.04$^{+  0.74}_{-  1.69  }$  &   1.53$^{+  0.06}_{-  0.16  }$  &   0.14$^{+  0.03}_{-  0.01  }$  &   8.63$^{+  0.52}_{-  1.11  }$  &   5.84 $\pm{  0.09  }$ \\
 A3395s   &   5.55$^{+  0.89}_{-  0.65  }$  &   2.12 $\pm{  0.13  }$  &   9.48$^{+  4.35}_{-  4.29  }$  &   1.78$^{+  0.24}_{-  0.32  }$  &   0.10$^{+  0.09}_{-  0.04  }$  &   9.39$^{+  2.49}_{-  3.88  }$  &   5.61 $\pm{  0.06  }$ \\
 A3526    &   3.69$^{+  0.05}_{-  0.04  }$  &   1.19 $\pm{  0.04  }$  &   3.41$^{+  0.60}_{-  1.36  }$  &   1.32$^{+  0.07}_{-  0.20  }$  &   0.11$^{+  0.04}_{-  0.01  }$  &   3.83$^{+  0.39}_{-  0.75  }$  &   3.66 $\pm{  0.16  }$ \\
 A3558    &   5.37$^{+  0.17}_{-  0.15  }$  &   6.56 $\pm{  0.04  }$  &   6.71$^{+  0.91}_{-  2.12  }$  &   1.59$^{+  0.07}_{-  0.19  }$  &   0.22$^{+  0.05}_{-  0.02  }$  &  14.60$^{+  0.84}_{-  2.35  }$  &  11.10 $\pm{  0.06  }$ \\
 A3562    &   4.47$^{+  0.23}_{-  0.21  }$  &   3.08 $\pm{  0.05  }$  &   3.51$^{+  0.43}_{-  0.91  }$  &   1.28$^{+  0.05}_{-  0.12  }$  &   0.20$^{+  0.04}_{-  0.02  }$  &   7.07$^{+  0.50}_{-  1.10  }$  &   3.23 $\pm{  0.08  }$ \\
 A3571    &   6.80$^{+  0.21}_{-  0.18  }$  &   8.08 $\pm{  0.11  }$  &   8.76$^{+  1.69}_{-  3.43  }$  &   1.75$^{+  0.10}_{-  0.27  }$  &   0.19$^{+  0.07}_{-  0.02  }$  &  16.45$^{+  1.36}_{-  2.93  }$  &  - \\
 A3581    &   1.83$^{+  0.04}_{-  0.02  }$  &   0.60 $\pm{  0.03  }$  &   0.93$^{+  0.19}_{-  0.38  }$  &   0.84$^{+  0.05}_{-  0.14  }$  &   0.15$^{+  0.07}_{-  0.03  }$  &   1.41$^{+  0.29}_{-  0.41  }$  &   1.22 $\pm{  0.09  }$ \\
 A3667    &   6.28$^{+  0.27}_{-  0.26  }$  &   9.48 $\pm{  0.11  }$  &   5.28$^{+  0.52}_{-  1.15  }$  &   1.46$^{+  0.05}_{-  0.11  }$  &   0.30$^{+  0.03}_{-  0.01  }$  &  16.04$^{+  1.15}_{-  2.22  }$  &   8.65 $\pm{  0.06  }$ \\
 A4038    &   3.22$^{+  0.10}_{-  0.10  }$  &   1.92 $\pm{  0.04  }$  &   2.58$^{+  0.49}_{-  1.05  }$  &   1.18$^{+  0.07}_{-  0.19  }$  &   0.16$^{+  0.08}_{-  0.03  }$  &   4.14$^{+  0.61}_{-  0.92  }$  &   2.85 $\pm{  0.06  }$ \\
 A4059    &   3.94$^{+  0.15}_{-  0.15  }$  &   2.80 $\pm{  0.06  }$  &   4.41$^{+  1.14}_{-  2.03  }$  &   1.39$^{+  0.11}_{-  0.26  }$  &   0.14$^{+  0.08}_{-  0.02  }$  &   6.17$^{+  0.79}_{-  1.40  }$  &   3.12 $\pm{  0.10  }$ \\
 COMA     &   8.07$^{+  0.29}_{-  0.27  }$  &   8.09 $\pm{  0.19  }$  &   9.95$^{+  2.10}_{-  2.99  }$  &   1.86$^{+  0.12}_{-  0.21  }$  &   0.19$^{+  0.07}_{-  0.04  }$  &  19.00$^{+  1.89}_{-  3.66  }$  &   8.94 $\pm{  0.05  }$ \\
 EXO0422  &   $^a$2.90$^{+  0.90}_{-  0.60  }$  &   2.03 $\pm{  0.21  }$  &   2.72$^{+  1.71}_{-  1.45  }$  &   1.19$^{+  0.21}_{-  0.27  }$  &   0.14$^{+  0.13}_{-  0.06  }$  &   3.68$^{+  2.04}_{-  1.75  }$  &   1.63 $\pm{  0.08  }$ \\
 FORNAX   &   1.56$^{+  0.05}_{-  0.07  }$  &   0.08 $\pm{  0.01  }$  &   1.29$^{+  0.44}_{-  0.55  }$  &   0.96$^{+  0.10}_{-  0.16  }$  &   0.04$^{+  0.02}_{-  0.01  }$  &   0.46$^{+  0.11}_{-  0.14  }$  &  - \\
 HYDRA-A  &   3.82$^{+  0.20}_{-  0.17  }$  &   5.84 $\pm{  0.04  }$  &   4.07$^{+  1.27}_{-  2.14  }$  &   1.34$^{+  0.13}_{-  0.29  }$  &   0.18$^{+  0.11}_{-  0.03  }$  &   7.29$^{+  0.77}_{-  1.72  }$  &   3.09 $\pm{  0.02  }$ \\
 IIIZw54  &  (3.00$^{+  1.39}_{-  0.95  }$) &   0.83 $\pm{  0.10  }$  &   3.76$^{+  2.82}_{-  2.26  }$  &   1.33$^{+  0.27}_{-  0.35  }$  &   0.08$^{+  0.12}_{-  0.05  }$  &   2.95$^{+  1.97}_{-  1.72  }$  &  - \\
 MKW3S    &   3.45$^{+  0.13}_{-  0.10  }$  &   2.79 $\pm{  0.05  }$  &   3.22$^{+  0.92}_{-  1.53  }$  &   1.25$^{+  0.11}_{-  0.24  }$  &   0.15$^{+  0.08}_{-  0.02  }$  &   4.92$^{+  0.59}_{-  1.06  }$  &   1.96 $\pm{  0.08  }$ \\
 MKW4     &   1.84$^{+  0.05}_{-  0.03  }$  &   0.34 $\pm{  0.01  }$  &   0.69$^{+  0.04}_{-  0.14  }$  &   0.76$^{+  0.01}_{-  0.06  }$  &   0.15$^{+  0.02}_{-  0.01  }$  &   1.05$^{+  0.07}_{-  0.15  }$  &   1.71 $\pm{  0.23  }$ \\
 MKW8     &   3.29$^{+  0.23}_{-  0.22  }$  &   0.79 $\pm{  0.11  }$  &   2.00$^{+  0.46}_{-  0.59  }$  &   1.08$^{+  0.08}_{-  0.12  }$  &   0.11$^{+  0.08}_{-  0.06  }$  &   2.22$^{+  1.11}_{-  1.29  }$  &   2.13 $\pm{  0.08  }$ \\
 NGC1550  &   1.44$^{+  0.03}_{-  0.02  }$  &   0.28 $\pm{  0.02  }$  &   0.68$^{+  0.14}_{-  0.24  }$  &   0.77$^{+  0.05}_{-  0.10  }$  &   0.12$^{+  0.07}_{-  0.04  }$  &   0.79$^{+  0.29}_{-  0.33  }$  &  - \\
 NGC4636  &   0.66$^{+  0.03}_{-  0.01  }$  &   0.02 $\pm{  0.00  }$  &   0.18$^{+  0.03}_{-  0.06  }$  &   0.49$^{+  0.03}_{-  0.07  }$  &   0.04$^{+  0.03}_{-  0.02  }$  &   0.08$^{+  0.04}_{-  0.04  }$  &  - \\
 NGC5044  &   1.22$^{+  0.04}_{-  0.04  }$  &   0.18 $\pm{  0.00  }$  &   0.49$^{+  0.12}_{-  0.25  }$  &   0.69$^{+  0.05}_{-  0.15  }$  &   0.09$^{+  0.04}_{-  0.01  }$  &   0.45$^{+  0.05}_{-  0.14  }$  &  - \\
 NGC507   &   1.40$^{+  0.04}_{-  0.07  }$  &   0.23 $\pm{  0.00  }$  &   0.46$^{+  0.02}_{-  0.07  }$  &   0.67$^{+  0.01}_{-  0.04  }$  &   0.15$^{+  0.02}_{-  0.01  }$  &   0.68$^{+  0.04}_{-  0.08  }$  &   1.94 $\pm{  0.16  }$ \\
 S1101    &   $^a$2.60$^{+  0.50}_{-  0.50  }$  &   3.52 $\pm{  0.05  }$  &   2.94$^{+  1.57}_{-  1.88  }$  &   1.20$^{+  0.18}_{-  0.34  }$  &   0.14$^{+  0.15}_{-  0.04  }$  &   4.22$^{+  0.90}_{-  1.23  }$  &   2.18 $\pm{  0.08  }$ \\
 ZwCl1215 &  (6.36$^{+  2.94}_{-  2.01  }$)  &   5.17 $\pm{  0.11  }$  &   9.46$^{+  5.74}_{-  4.87  }$  &   1.74$^{+  0.30}_{-  0.37  }$  &   0.15$^{+  0.08}_{-  0.04  }$  &  13.87$^{+  3.01}_{-  3.76  }$  &  - \\
\hline\noalign{Note: The values of $T_h$ and $L_X$ are from Ikebe et al. (2002). 
$T_h$ in a bracket is estimated from the $L_X-T$ relation (Ikebe et al. 2002). 
$T_m$ with `a' is derived from non-ASCA spectroscopy (Ikebe et al. 2002 and references therein). }
\end{tabular}
\end{table*}

\begin{table*}
\renewcommand{\arraystretch}{0.9}
\caption[]{Cluster properties for the scaling relations of the extended cluster sample.}
\begin{tabular}{ccccccccc}
\hline
\hline\noalign{}
\hline\noalign{}
 Name  & $T_h$  & $L_X$  & $M_{500}$  & $r_{500}$  & $f_{gas}$
 & $M_{gas,500}$  & $L_{nir}$ \\
    & (keV)  & ($10^{44}$ erg/s, 0.1-2.4keV) &  ($10^{14}$
 M$_{\sun}$) & (Mpc) &   &  ($10^{13}$ M$_{\sun}$)
  & ($h_{70}^{-2} \, 10^{12}$ L$_{\sun}$) \\
\hline\noalign{}
 3C129    &   5.57$^{+  0.16}_{-  0.15  }$  &   2.27 $\pm{  0.21  }$  &   5.39$^{+  2.26}_{-  2.33  }$  &   1.51$^{+  0.19}_{-  0.26  }$  &   0.16$^{+  0.16}_{-  0.10  }$  &   8.55$^{+  2.58}_{-  4.82  }$  &  - \\
 A0539    &   3.04$^{+  0.11}_{-  0.10  }$  &   1.11 $\pm{  0.02  }$  &   2.68$^{+  0.32}_{-  0.85  }$  &   1.19$^{+  0.05}_{-  0.14  }$  &   0.14$^{+  0.05}_{-  0.01  }$  &   3.78$^{+  0.27}_{-  0.47  }$  &   3.72 $\pm{  0.05  }$ \\
 A0548e   &   2.93$^{+  0.17}_{-  0.15  }$  &   1.05 $\pm{  0.03  }$  &   1.49$^{+  0.13}_{-  0.17  }$  &   0.97$^{+  0.03}_{-  0.04  }$  &   0.18$^{+  0.02}_{-  0.02  }$  &   2.65$^{+  0.27}_{-  0.29  }$  &   4.17 $\pm{  0.08  }$ \\
 A0548w   &  (1.68$^{+  0.77}_{-  0.53  }$) &   0.19 $\pm{  0.02  }$  &   1.00$^{+  0.58}_{-  0.52  }$  &   0.85$^{+  0.14}_{-  0.19  }$  &   0.08$^{+  0.07}_{-  0.04  }$  &   0.82$^{+  0.45}_{-  0.43  }$  &  - \\
 A0644    &   6.54$^{+  0.27}_{-  0.26  }$  &   8.35 $\pm{  0.15  }$  &   8.41$^{+  2.15}_{-  3.81  }$  &   1.68$^{+  0.13}_{-  0.31  }$  &   0.16$^{+  0.07}_{-  0.02  }$  &  13.85$^{+  1.56}_{-  2.97  }$  &   6.34 $\pm{  0.06  }$ \\
 A1413    &   6.56$^{+  0.65}_{-  0.44  }$  &  10.71 $\pm{  0.30  }$  &   9.77$^{+  2.78}_{-  4.58  }$  &   1.65$^{+  0.14}_{-  0.31  }$  &   0.17$^{+  0.09}_{-  0.03  }$  &  16.85$^{+  1.94}_{-  3.46  }$  &  - \\
 A1689    &   8.58$^{+  0.84}_{-  0.40  }$  &  19.48 $\pm{  0.34  }$  &  14.98$^{+  5.82}_{-  8.38  }$  &   1.84$^{+  0.21}_{-  0.44  }$  &   0.16$^{+  0.13}_{-  0.03  }$  &  23.26$^{+  3.69}_{-  5.42  }$  &  - \\
 A1775    &   3.66$^{+  0.34}_{-  0.20  }$  &   3.09 $\pm{  0.09  }$  &   4.19$^{+  1.36}_{-  1.69  }$  &   1.32$^{+  0.13}_{-  0.21  }$  &   0.18$^{+  0.07}_{-  0.03  }$  &   7.55$^{+  2.21}_{-  2.07  }$  &  - \\
 A1800    &  (5.02$^{+  2.32}_{-  1.59  }$) &   2.85 $\pm{  0.37  }$  &   5.94$^{+  4.83}_{-  3.59  }$  &   1.49$^{+  0.33}_{-  0.40  }$  &   0.13$^{+  0.16}_{-  0.09  }$  &   7.63$^{+  4.25}_{-  4.68  }$  &  - \\
 A1914    &   8.41$^{+  0.60}_{-  0.58  }$  &  17.04 $\pm{  0.38  }$  &  11.84$^{+  3.65}_{-  5.84  }$  &   1.72$^{+  0.16}_{-  0.35  }$  &   0.18$^{+  0.10}_{-  0.03  }$  &  21.20$^{+  2.54}_{-  4.72  }$  &  - \\
 A2151w   &   2.58$^{+  0.19}_{-  0.20  }$  &   0.89 $\pm{  0.03  }$  &   1.60$^{+  0.35}_{-  0.61  }$  &   1.00$^{+  0.07}_{-  0.15  }$  &   0.14$^{+  0.05}_{-  0.02  }$  &   2.25$^{+  0.24}_{-  0.51  }$  &   3.44 $\pm{  0.14  }$ \\
 A2319    &   8.84$^{+  0.29}_{-  0.24  }$  &  16.37 $\pm{  0.26  }$  &  13.57$^{+  2.15}_{-  4.59  }$  &   1.99$^{+  0.10}_{-  0.26  }$  &   0.24$^{+  0.07}_{-  0.02  }$  &  32.26$^{+  3.88}_{-  5.92  }$  &  14.63 $\pm{  0.04  }$ \\
 A2734    &   5.07$^{+  0.36}_{-  0.42  }$  &   2.40 $\pm{  0.10  }$  &   4.82$^{+  0.85}_{-  1.56  }$  &   1.40$^{+  0.08}_{-  0.17  }$  &   0.14$^{+  0.05}_{-  0.02  }$  &   6.80$^{+  0.97}_{-  1.52  }$  &   3.13 $\pm{  0.09  }$ \\
 A2877    &   $^a$3.50$^{+  2.20}_{-  1.10  }$  &   0.40 $\pm{  0.01  }$  &   6.88$^{+  6.74}_{-  3.79  }$  &   1.64$^{+  0.42}_{-  0.38  }$  &   0.04$^{+  0.03}_{-  0.02  }$  &   3.04$^{+  0.51}_{-  0.63  }$  &  - \\
 A3395n   &   5.11$^{+  0.47}_{-  0.43  }$  &   1.63 $\pm{  0.11  }$  &   8.13$^{+  5.53}_{-  4.69  }$  &   1.69$^{+  0.32}_{-  0.42  }$  &   0.10$^{+  0.12}_{-  0.07  }$  &   8.24$^{+  2.20}_{-  5.34  }$  &  - \\
 A3528n   &   4.79$^{+  0.50}_{-  0.44  }$  &   1.56 $\pm{  0.06  }$  &   4.49$^{+  0.78}_{-  1.53  }$  &   1.38$^{+  0.08}_{-  0.18  }$  &   0.12$^{+  0.04}_{-  0.02  }$  &   5.33$^{+  0.67}_{-  1.18  }$  &  - \\
 A3528s   &   4.60$^{+  0.49}_{-  0.27  }$  &   2.20 $\pm{  0.06  }$  &   2.76$^{+  0.39}_{-  0.44  }$  &   1.17$^{+  0.05}_{-  0.07  }$  &   0.20$^{+  0.03}_{-  0.02  }$  &   5.56$^{+  0.51}_{-  0.73  }$  &  - \\
 A3530    &   4.05$^{+  0.32}_{-  0.30  }$  &   1.21 $\pm{  0.06  }$  &   4.34$^{+  1.09}_{-  1.27  }$  &   1.37$^{+  0.11}_{-  0.15  }$  &   0.12$^{+  0.05}_{-  0.04  }$  &   5.22$^{+  0.89}_{-  1.38  }$  &  - \\
 A3532    &   4.41$^{+  0.19}_{-  0.18  }$  &   2.20 $\pm{  0.06  }$  &   6.63$^{+  1.17}_{-  2.84  }$  &   1.57$^{+  0.09}_{-  0.27  }$  &   0.12$^{+  0.08}_{-  0.01  }$  &   8.11$^{+  0.68}_{-  1.38  }$  &  - \\
 A3560    &  (3.90$^{+  1.81}_{-  1.23  }$) &   1.57 $\pm{  0.06  }$  &   2.77$^{+  1.85}_{-  1.26  }$  &   1.18$^{+  0.22}_{-  0.22  }$  &   0.17$^{+  0.06}_{-  0.04  }$  &   4.61$^{+  1.52}_{-  1.36  }$  &   6.50 $\pm{  0.27  }$ \\
 A3627    &   5.62$^{+  0.12}_{-  0.11  }$  &   3.59 $\pm{  0.18  }$  &   4.92$^{+  0.67}_{-  0.91  }$  &   1.48$^{+  0.06}_{-  0.10  }$  &   0.19$^{+  0.05}_{-  0.04  }$  &   9.57$^{+  1.37}_{-  2.14  }$  &  - \\
 A3695    &  (6.76$^{+  3.12}_{-  2.14  }$) &   5.89 $\pm{  0.89  }$  &   7.03$^{+  4.66}_{-  4.22  }$  &   1.55$^{+  0.29}_{-  0.41  }$  &   0.18$^{+  0.18}_{-  0.12  }$  &  12.82$^{+  6.66}_{-  7.84  }$  &  - \\
 A3822    &   5.12$^{+  0.43}_{-  0.31  }$  &   4.83 $\pm{  0.58  }$  &   4.69$^{+  1.19}_{-  1.46  }$  &   1.37$^{+  0.11}_{-  0.16  }$  &   0.19$^{+  0.13}_{-  0.08  }$  &   9.07$^{+  2.53}_{-  4.06  }$  &   5.38 $\pm{  0.10  }$ \\
 A3827    &  (7.66$^{+  3.54}_{-  2.42  }$) &   7.94 $\pm{  0.78  }$  &  15.50$^{+ 13.81}_{-  9.76  }$  &   2.01$^{+  0.48}_{-  0.56  }$  &   0.13$^{+  0.19}_{-  0.08  }$  &  19.88$^{+ 11.11}_{- 11.45  }$  &  - \\
 A3888    &  (8.68$^{+  4.01}_{-  2.75  }$) &  10.09 $\pm{  0.39  }$  &  29.81$^{+ 24.98}_{- 21.73  }$  &   2.38$^{+  0.54}_{-  0.84  }$  &   0.07$^{+  0.13}_{-  0.03  }$  &  21.97$^{+  3.60}_{-  6.63  }$  &  - \\
 A3921    &   5.39$^{+  0.38}_{-  0.35  }$  &   4.83 $\pm{  0.14  }$  &   6.59$^{+  1.50}_{-  2.32  }$  &   1.51$^{+  0.11}_{-  0.20  }$  &   0.16$^{+  0.06}_{-  0.03  }$  &  10.76$^{+  1.41}_{-  2.05  }$  &  - \\
 AWM7     &   3.70$^{+  0.08}_{-  0.05  }$  &   2.10 $\pm{  0.07  }$  &   4.92$^{+  1.21}_{-  2.26  }$  &   1.48$^{+  0.11}_{-  0.27  }$  &   0.12$^{+  0.07}_{-  0.02  }$  &   5.85$^{+  0.70}_{-  1.10  }$  &   2.90 $\pm{  0.06  }$ \\
 HCG94    &   3.30$^{+  0.17}_{-  0.16  }$  &   1.28 $\pm{  0.02  }$  &   2.25$^{+  0.24}_{-  0.60  }$  &   1.11$^{+  0.04}_{-  0.11  }$  &   0.15$^{+  0.03}_{-  0.01  }$  &   3.41$^{+  0.18}_{-  0.49  }$  &   2.51 $\pm{  0.11  }$ \\
 IIZw108  &  (4.28$^{+  1.98}_{-  1.35  }$) &   1.98 $\pm{  0.24  }$  &   3.85$^{+  2.60}_{-  2.00  }$  &   1.32$^{+  0.25}_{-  0.28  }$  &   0.15$^{+  0.11}_{-  0.07  }$  &   5.65$^{+  2.88}_{-  2.73  }$  &  - \\
 M49      &   1.33$^{+  0.03}_{-  0.03  }$  &   0.02 $\pm{  0.00  }$  &   0.67$^{+  0.40}_{-  0.47  }$  &   0.77$^{+  0.13}_{-  0.26  }$  &   0.01$^{+  0.01}_{-  0.00  }$  &   0.09$^{+  0.02}_{-  0.04  }$  &  - \\
 NGC499   &   0.66$^{+  0.02}_{-  0.03  }$  &   0.04 $\pm{  0.00  }$  &   0.33$^{+  0.25}_{-  0.23  }$  &   0.60$^{+  0.12}_{-  0.20  }$  &   0.03$^{+  0.04}_{-  0.01  }$  &   0.08$^{+  0.02}_{-  0.03  }$  &  - \\
 NGC5813  &   0.76$^{+  0.19}_{-  0.19  }$  &   0.02 $\pm{  0.00  }$  &   0.43$^{+  0.45}_{-  0.33  }$  &   0.66$^{+  0.18}_{-  0.26  }$  &   0.01$^{+  0.05}_{-  0.01  }$  &   0.06$^{+  0.07}_{-  0.04  }$  &  - \\
 NGc5846  &   0.64$^{+  0.04}_{-  0.03  }$  &   0.01 $\pm{  0.00  }$  &   0.18$^{+  0.11}_{-  0.12  }$  &   0.49$^{+  0.08}_{-  0.16  }$  &   0.03$^{+  0.03}_{-  0.01  }$  &   0.05$^{+  0.02}_{-  0.02  }$  &  - \\
 OPHIUCHU &  10.25$^{+  0.30}_{-  0.36  }$  &  12.14 $\pm{  0.39  }$  &  38.76$^{+ 21.59}_{- 24.11  }$  &   2.91$^{+  0.46}_{-  0.81  }$  &   0.07$^{+  0.09}_{-  0.02  }$  &  28.60$^{+  2.89}_{-  6.11  }$  &  - \\
 PERSEUS  &   6.42$^{+  0.06}_{-  0.06  }$  &  16.39 $\pm{  0.20  }$  &   6.08$^{+  1.55}_{-  2.85  }$  &   1.58$^{+  0.12}_{-  0.30  }$  &   0.30$^{+  0.12}_{-  0.03  }$  &  18.10$^{+  1.81}_{-  4.79  }$  &   8.63 $\pm{  0.07  }$ \\
 PKS0745  &   6.37$^{+  0.21}_{-  0.20  }$  &  27.13 $\pm{  0.42  }$  &   7.11$^{+  2.85}_{-  3.91  }$  &   1.54$^{+  0.18}_{-  0.36  }$  &   0.28$^{+  0.18}_{-  0.05  }$  &  19.74$^{+  2.46}_{-  5.35  }$  &  - \\
 RXJ2344  &  (5.52$^{+  2.55}_{-  1.74  }$) &   3.59 $\pm{  0.08  }$  &   8.89$^{+  5.48}_{-  5.25  }$  &   1.70$^{+  0.29}_{-  0.44  }$  &   0.11$^{+  0.10}_{-  0.03  }$  &   9.47$^{+  1.70}_{-  2.29  }$  &  - \\
 S405     &  (5.02$^{+  2.32}_{-  1.59  }$) &   2.90 $\pm{  0.39  }$  &   4.62$^{+  3.14}_{-  2.88  }$  &   1.39$^{+  0.26}_{-  0.39  }$  &   0.16$^{+  0.13}_{-  0.10  }$  &   7.59$^{+  3.71}_{-  4.37  }$  &  - \\
 S540     &  (3.09$^{+  1.43}_{-  0.98  }$) &   0.89 $\pm{  0.07  }$  &   2.52$^{+  1.55}_{-  1.40  }$  &   1.16$^{+  0.20}_{-  0.27  }$  &   0.11$^{+  0.08}_{-  0.05  }$  &   2.68$^{+  1.18}_{-  1.23  }$  &  - \\
 S636     &   2.06$^{+  0.07}_{-  0.06  }$  &   0.38 $\pm{  0.03  }$  &   1.56$^{+  0.44}_{-  0.47  }$  &   1.01$^{+  0.09}_{-  0.11  }$  &   0.10$^{+  0.07}_{-  0.04  }$  &   1.52$^{+  0.35}_{-  0.62  }$  &  - \\
 TRIANGUL &   9.06$^{+  0.33}_{-  0.31  }$  &  12.43 $\pm{  0.15  }$  &  14.84$^{+  2.49}_{-  5.28  }$  &   2.07$^{+  0.11}_{-  0.28  }$  &   0.20$^{+  0.06}_{-  0.02  }$  &  29.14$^{+  2.35}_{-  4.92  }$  &  12.01 $\pm{  0.05  }$ \\
 UGC03957 &  (3.21$^{+  1.48}_{-  1.02  }$) &   0.98 $\pm{  0.10  }$  &   3.32$^{+  2.76}_{-  2.03  }$  &   1.28$^{+  0.28}_{-  0.34  }$  &   0.08$^{+  0.10}_{-  0.04  }$  &   2.67$^{+  1.62}_{-  1.47  }$  &  - \\
 ZwCl1742 &  (6.05$^{+  2.80}_{-  1.91  }$) &   4.54 $\pm{  0.33  }$  &  10.11$^{+  5.91}_{-  6.30  }$  &   1.78$^{+  0.30}_{-  0.49  }$  &   0.11$^{+  0.13}_{-  0.03  }$  &  11.20$^{+  3.42}_{-  3.22  }$  &  - \\
\hline\noalign{Note: The symbols in $T_h$ have the same meanings as in Table 4.}
\end{tabular}
\end{table*}

\subsection{The $M$--$T_h$ relation}

Figure 5 shows the relation of the globally measured X-ray temperature
of the hot component, $T_h$ (excluding  non-ASCA derived $T_h$), 
and the
cluster mass, $M_{500}$. The self-similar model prediction is a slope with a
value of 1.5, consistent with the value we obtain for the total sample with
measured $T_h$ and
consistent with the discussion in Finoguenov et al. (2001).
The results of power-law fits to all the relations discussed in this section
are summarized in Table 7. Moreover, the slopes and the normalizations of the 
$M$--$T_h$
relation for the CCCs and NCCCs are consistent within errors. Thus there is no
significant influence of cooling cores on this relation. Note, however, that
the two cluster parameters compared in this relation, $T_h$ and $M_{500}$, are
not independently obtained, but $M_{500}$ is directly dependent on the
temperature measurements. If $T_h$ has an offset $\Delta T_h$, $M_{500}$
will change to $M_{500}+\Delta M_{500} \propto 
(T_h+\Delta T_h)^{3/2}$. Note that
this slope is the same as the self-similar model prediction.
Thus the cooling cores' influence on temperature determination will appear
in the cluster mass such
that the overall effect may remain undetected (see also discussion below).

Among the relations listed in Table 7, in conjunction with the $M_{500}$--$T_h$ relation,
together with the $M_{gas,500}$--$T_h$ relation (Fig. 11), 
has the smallest scatter.
This could be due in part to the correlation of the mass and temperature parameter. 
We use the $M_{gas,500}$--$T_h$ relation to show below (Sect. 4.5) 
that this is not a strong effect and thus not the main reason
for the different $L_X$ normalization. Instead, the tight relation shows
that mass and temperature
are linked in a more fundamental way by simple self-similar gravitational
processes than by the other relations.

\subsection{The $L_X$--$T$ relation}

The $L_X$--$T_m$ and $L_X$--$T_h$ relations for the cluster sample with
measured $T_m$ and $T_h$ derived from ASCA are
shown in Figs. 6 and 7. The values of $L_X$ are the 
X-ray luminosities in (0.1-2.4 keV) derived from Ikebe et al. (2002).
Since for some clusters $T_m$ is measured with the central region 
excluded (cooling flow correction),
here we only use $T_m$ without any cooling flow correction.
For $T_m$ derived from Fukazawa et al. (1998), which includes cooling flow
correction, we use the central $T_m$ (0-$2\arcmin \sim 3\arcmin$) instead 
(Fukazawa et al. 2000).  
The resulting slopes of $L_X$--$T$ relations are $2.23\pm0.15$ and $2.73\pm0.13$
for $T_m$ and $T_h$, respectively.
They are much higher than the 1.5 predicted by a self-similar model 
for the $L_X$--$T$ relation (note $L_X$ is in the ROSAT band, not bolometric).
This is consistent with the results in Reiprich \& B\"ohringer (2002).
Note that the slope of $L_X$--$T_m$ is shallower than that of $L_X$--$T_h$.
This may be due to $T_m$ having an offset to the low
temperature direction compared to $T_h$ and fewer low $L_X$ clusters
being included in the $T_m$ relation.  
Remarkable is the clearly higher normalization of the relation for CCCs
compared to NCCCs, with offsets of factors of 2.05 and 1.84 for the relation with
$T_m$ and $T_h$, respectively. The normalization difference is slightly
small using $T_h$ in the scaling relation and the scatter is also slightly reduced, since
$T_h$ most probably provides a better measure of the global gravitational
potential depth and better mass proxy, as this temperature is not as 
downward-biased
by the central cooling core region as $T_m$. But the difference between
the two relations is not very large. This is already an indication that biased
temperature measurements for CCCs are not the major reason for the different 
normalizations
of the relations for CCCs and NCCCs. Here the effect of the different mass
coverage of CCCs and NCCCs on the slope of the combined relation (which should
make the relation shallower) is not as strong as in Fig. 4. 
The combined relation is
slightly shallower than the NCCC and CCC relations.
 
\subsection{The $L_X$--$M$ relation}

The $L_X$--$M$ relation is the most important relation for the application
to cosmological cluster surveys. The previously determined relation
for this sample (Reiprich \& B\"ohringer, 2002) was used to get cluster mass
estimates, for example, for the cosmological studies (Schuecker et al. 2001b, 2003;
Stanek et al. 2006). The resulting slope of $L_X$--$M$ 
relation is again higher
than 1.0, which is predicted by a self-similar model, as shown in Fig. 8.

Like in the X-ray luminosity temperature relation,
we see a substantial difference in the normalization of this relation
for the CCCs and NCCCs by about a factor of 2.4. The offset between the 
different cluster subsamples
shows that the scatter in the overall sample is partly produced by the
different types of clusters, and knowing more about the clusters 
helps reduce this scatter as discussed below.

One important question concerns the origin of this large difference.
There are, in principle, two effects caused by cool cores that add to 
the observed result: cluster temperatures and the cluster masses derived 
from temperature estimates will be biased low and luminosities will be 
biased high due to the enhanced emission of dense cores and more compact 
clusters. In turn, if the core radii of the
NCCCs are inflated, the cluster masses will be biased high for the NCCCs.

To distinguish these different possibilities, we need mass estimators independent
of the temperature and independent of the core radii. In the next section
we apply these new parameters. 
Figure 8 also shows that the fraction of CCCs at the high $L_X$ end
do not constitute a small fraction of all clusters unlike 
that at the high $M_{500}$ end. 
This is due to CCCs usually having higher $L_X$ for
the clusters with the same $M_{500}$.
The CCC fraction is about 60\% (11 CCCs) for the 
most luminous 18 clusters (with $L_X > $ 6 10$^{44}$ erg/s 
and $z <0.17$ in the $\Lambda$CDM cosmology) in this sample.
This fraction is consistent with the distant cluster 
samples with similar $L_X$, 
e.g., 7 CCCs out of 12 clusters at $z \sim 0.2$ (Zhang et al. 2006b) and
6 CCCs out of 13 clusters at $z \sim 0.3$ (Zhang et al. 2006a), where
we use the same criterion of CCCs as in the HIFLUCS.
This result shows that the fraction of CCCs in luminous cluster samples
does not show a large evolutionary effect up to $z \sim 0.3$.

\subsection{Relations involving total NIR luminosity and gas mass}

Assuming that the cluster gas mass fraction is approximately constant with cluster
mass (e.g. Allen et al. 2004; Ettori et al. 2003), we can also use the cluster
gas mass to estimate the cluster's total mass. The determination of the
cluster gas mass depends only on the X-ray surface-brightness distribution, 
not on the temperature. However, the way we determine 
$M_{gas,500}$, based on the fiducial outer radius of $r_{500}$,
introduces a weak temperature dependence, since
in our approach we used the temperature-dependent gravitational mass
to estimate $r_{500}$. This dependence is roughly proportional to
$T^{1/2}$ and thus much weaker than proportional
and much weaker than the dependence of $M_{500} \propto T^{3/2}$.
Thus if there is any strong bias in the temperature determination due to
cooling flows, we should still see this effect in a correlation analysis based on
gas mass, but it will just be weakened approximately by a factor of two.
For example, if the temperature of a cluster, $T$, is biased to 2$T$,
the measured $M_{gas,500}$ will change to 1.4$M_{gas,500}$. However, if this
cluster remains on the line with a slope 2.0 (see Table 7), it needs 
$M_{gas,500}$ to change to 4$M_{gas,500}$. From this example, one can
see that $M_{gas,500}$ is insensitive to $T$ in the relation $M_{gas,500}$--$T$. 
Another mass estimator is the total luminosity of the cluster
galaxies in the NIR (K-band), which is obtained from the
Two Micron All Sky Survey (2MASS) (Lin et al. 2004).

The $L_{nir}$--$T_h$ relation in Fig. 9 shows no strong bias of CCCs versus NCCCs.
The normalization difference is less than 10\% so well within the 
$1\sigma$ error
of the fits. The comparison of the two mass estimators $L_{nir}$ and
$M_{gas,500}$ in Fig. 10 shows a
comparatively small difference between the CCCs and NCCCs with a factor smaller than
about 1.25, where the CCCs have on average a little higher $M_{gas,500}$.
Therefore we do not see a strong bias in either of the two mass estimators,
although a weak bias cannot be ruled out, especially in the gas mass.

Similarly, in Figs. 11 and 12 where we show the $M_{gas,500}$--$T_h$ and
$M_{gas,500}$--$T_m$ relations, we do not see a strong temperature bias for CCCs
versus NCCCs. Similar to Figs. 6 and 7, the slope of $M_{gas,500}$--$T_m$
is shallower than that of $M_{gas,500}$--$T_h$, but it only has a slight
difference for the clusters with $T_h > 3$keV. Note that
clusters with $T_m<1.0$keV in Fig. 11 are not plotted in Fig. 12.
In addition, the difference in the normalization of this relation
for the CCCs and NCCCs is very small and
within the measurement errors, with a factor 1.10 and 1.00 for $T_m$
and $T_h$, respectively.
Therefore we conclude from the results in this section that the segregation
of CCCs and NCCCs in the $L_X$--$M$ and $L_X$--$T$ relations is mainly an
X-ray luminosity effect and, to a lesser extent,
an effect of a biased temperature estimate.
A similar conclusion has been reached by O'Hara et al. (2006).

\subsection{Gas mass fraction}

We found no difference of the total gas mass fraction, $f_{gas,500}$, 
between CCCs and NCCCs
as shown in Fig. 13. This reconfirms the weak influence of cooling cores
on the mass and gas mass estimates.

\begin{table*}
\renewcommand{\arraystretch}{1.0}
\caption[]{Fits to relations of $r_c$ and $M_{500}$.}
\begin{tabular}{ccccc}
\hline
\hline\noalign{}
 relation  & Number of clusters & $B$ & $A$  & comments
  \\
\hline\noalign{}
$r_{c}-M_{500}$   & 106    & $1.18 \pm{0.08}$ & $0.253 \pm{0.036}$ & ALL  \\
$r_{c}-M_{500}$   & 52     & $0.90 \pm{0.06}$ & $0.023 \pm{0.036}$ & CCCs \\
$r_{c}-M_{500}$   & 54     & $0.66 \pm{0.12}$ & $0.452 \pm{0.029}$ & NCCCs \\
\hline\noalign{Note: The relations are given in the form:
$ \log_{10}\left({r_{c} \over 100  {\rm kpc}}\right) = A
+ B\cdot \log_{10} \left({M_{500} \over 5 10^{14}{\rm M_{\odot}}}\right)$.}
\end{tabular}
\end{table*}

\begin{table*}
\renewcommand{\arraystretch}{1.0}
\caption[]{Summary of the fits to the scaling relations. }
\begin{tabular}{ccccc}
\hline
\hline\noalign{}
 relation  & Number of clusters & $B$ & $A$  & comments
  \\
\hline\noalign{}
$M_{500}$--$T_h$   & 88    & $1.54 \pm{0.06}$ & $-0.112 \pm{0.014}$ & ALL with $T_h$ \\
$M_{500}$--$T_h$   & 47    & $1.48 \pm{0.07}$ & $-0.140 \pm{0.015}$ & CCCs \\
$M_{500}$--$T_h$   & 41    & $1.57 \pm{0.17}$ & $-0.088 \pm{0.031}$ & NCCCs \\
$M_{500}$--$T_h$   & 72    & $1.30 \pm{0.18}$ & $-0.069 \pm{0.025}$ & $T_h>3.0$ keV \\
$M_{500}$--$T_h$   & 16    & $0.91 \pm{0.31}$ & $-0.46 \pm{0.13}$ & $T_h<3.0$ keV \\
$M_{gas,500}$--$T_h$   & 88    & $2.29 \pm{0.09}$ & $-0.269 \pm{0.015}$ & ALL with $T_h$ \\
$M_{gas,500}$--$T_h$   & 47    & $2.38 \pm{0.10}$ & $-0.251 \pm{0.017}$ & CCCs \\
$M_{gas,500}$--$T_h$   & 41    & $2.04 \pm{0.14}$ & $-0.258 \pm{0.025}$ & NCCCs \\
$M_{gas,500}$--$T_h$   & 72    & $1.89 \pm{0.10}$ & $-0.221 \pm{0.017}$ & $T_h>3.0$ keV \\
$M_{gas,500}$--$T_h$   & 35    & $1.98 \pm{0.11}$ & $-0.219 \pm{0.019}$ & $T_h>3.0$ keV, CCCs \\
$M_{gas,500}$--$T_h$   & 37    & $1.80 \pm{0.18}$ & $-0.219 \pm{0.033}$ & $T_h>3.0$ keV, NCCCs \\
$M_{gas,500}$--$T_m$   & 71    & $1.70 \pm{0.06}$ & $-0.184 \pm{0.015}$ & ALL with $T_m$ \\
$M_{gas,500}$--$T_m$   & 38    & $1.73 \pm{0.07}$ & $-0.166 \pm{0.022}$ & CCCs \\
$M_{gas,500}$--$T_m$   & 33    & $1.71 \pm{0.11}$ & $-0.207 \pm{0.025}$ & NCCCs \\
$L_X$--$T_h$       & 88    & $2.73 \pm{0.13}$ & $0.363 \pm{0.027}$ & ALL with $T_h$ \\
$L_X$--$T_h$       & 47    & $2.88 \pm{0.15}$ & $0.492 \pm{0.031}$ & CCCs \\
$L_X$--$T_h$       & 41    & $2.74 \pm{0.17}$ & $0.227 \pm{0.034}$ & NCCCs  \\
$L_X$--$T_h$       & 72    & $2.88 \pm{0.19}$ & $0.332 \pm{0.034}$ & $T_h>3.0$ keV  \\
$L_X$--$T_h$       & 35    & $3.08 \pm{0.24}$ & $0.457 \pm{0.039}$ & $T_h>3.0$ keV, CCCs  \\
$L_X$--$T_h$       & 37    & $2.78 \pm{0.22}$ & $0.213 \pm{0.044}$ & $T_h>3.0$ keV, NCCCs \\
$L_X$--$T_m$       & 71    & $2.23 \pm{0.11}$ & $0.458 \pm{0.032}$ & ALL with $T_m$ \\
$L_X$--$T_m$       & 38    & $2.31 \pm{0.14}$ & $0.597 \pm{0.040}$ & CCCs \\
$L_X$--$T_m$       & 33    & $2.33 \pm{0.09}$ & $0.286 \pm{0.031}$ & NCCCs  \\
$L_X$--$M_{500}$ & 106   & $1.82 \pm{0.13}$ & $0.521 \pm{0.039}$  & ALL  \\
$L_X$--$M_{500}$ & 88    & $1.77 \pm{0.12}$ & $0.562 \pm{0.041}$  & ALL with $T_h$ \\
$L_X$--$M_{500}$ & 47    & $1.94 \pm{0.15}$ & $0.763 \pm{0.050}$  & CCCs \\
$L_X$--$M_{500}$ & 41    & $1.75 \pm{0.25}$ & $0.381 \pm{0.062}$  & NCCCs \\
$L_X$--$M_{500}$ & 72    & $2.23 \pm{0.38}$ & $0.485 \pm{0.053}$  & $T_h>3.0$ keV \\
$L_X$--$M_{500}$ & 35    & $3.05 \pm{0.60}$ & $0.674 \pm{0.074}$  & $T_h>3.0$ keV, CCCs \\
$L_X$--$M_{500}$ & 37    & $1.84 \pm{0.42}$ & $0.350 \pm{0.087}$  & $T_h>3.0$ keV, NCCCs \\
$L_{nir}$--$T_h$       & 58    & $1.34 \pm{0.09}$ & $0.586 \pm{0.017}$ & ALL with $T_m$ \\
$L_{nir}$--$T_h$       & 31    & $1.17 \pm{0.11}$ & $0.565 \pm{0.017}$ & CCCs \\
$L_{nir}$--$T_h$       & 27    & $1.47 \pm{0.15}$ & $0.593 \pm{0.035}$ & NCCCs  \\
$L_{nir}$--$M_{gas,500}$   & 62    & $0.702 \pm{0.055}$ & $0.741 \pm{0.016}$ & ALL  \\
$L_{nir}$--$M_{gas,500}$   & 33    & $0.604 \pm{0.062}$ & $0.686 \pm{0.018}$ & CCCs \\
$L_{nir}$--$M_{gas,500}$   & 29    & $0.756 \pm{0.080}$ & $0.778 \pm{0.024}$ & NCCCs  \\
\hline\noalign{Note: The relations are given in the form:
$\log_{10}\left({M_{500} \over 5\, 10^{14} {\rm M}_{\odot}}\right) = A
+ B\cdot \log_{10} \left({T \over 4{\rm keV}}\right)$,
$\log_{10}\left({M_{gas,500} \over 10^{14} {\rm M}_{\odot}}\right) = A
+ B\cdot \log_{10} \left({T \over 4{\rm keV}}\right)$,
$\log_{10}\left({L_X \over  10^{44} {\rm erg/s}}\right) = A
+ B\cdot \log_{10} \left({T \over 4{\rm keV}}\right)$,
$\log_{10}\left({L_X \over  10^{44} {\rm erg/s}}\right) = A
+ B\cdot \log_{10} \left({M_{500} \over 5\, 10^{14} {\rm M}_{\odot}}\right)$,
$\log_{10}\left({L_{nir}}\right) = A
+ B\cdot \log_{10} \left({T \over 4{\rm keV}}\right)$ and
$\log_{10}\left({L_{nir}}\right) = A
+ B\cdot \log_{10} \left({M_{gas,500} \over 10^{14} {\rm M}_{\odot}}\right)$.}
\end{tabular}
\end{table*}

\section{Discussions and summary}

In this paper we have used an isothermal model to determine the X-ray mass, 
because to date only global temperature estimates mainly 
from ASCA are available for such a large sample. 
For the central region in cooling core clusters, this is obviously not correct.
But we expect that the total gravitational mass is correct at large radii, 
which is confirmed by our mass determination for the cluster 
PKS 0745-191 and Abell 1650 based on
XMM-Newton observations (Chen et al. 2003, Jia et al. 2006)
in which the resultant total masses
are found to be consistent with this work. 

In summary we find from the analysis presented in this paper that:

(i) the formally-derived mass deposition rates for the strongest cooling 
core clusters are roughly proportional to the cluster mass.

(ii) the fraction of NCCCs increases significantly with $M_{500}$, and
most of the galaxy groups in HIFLUGCS are cooling core clusters. 
This is most probably explained by the fact that the most massive 
galaxy clusters have
been formed more recently than the others and should therefore show a larger 
fraction of dynamically young systems. These may turn into cooling core
clusters in a later evolutionary stage. In addition, the fraction of 
CCCs in luminous cluster samples  
does not show a large evolution effect within $z < 0.3$.

(iii) among all the observational parameters, the core radius and the X-ray 
luminosity are shown to be most affected by the presence of a cooling core, as 
observed in their relation
to other bulk cluster properties as cluster mass and temperature.

(iv) the $M$--$T$ relation using the X-ray temperature of the hot ICM phase,
$T_h$, seems to show a comparatively small bias for
CCCs in comparison to NCCCs.

From the magnitude of the effect (iii) we can conclude that the scatter in the
$L_X - M_{500}$ relation, which is so important in cosmological applications, is
to a large part due to the different normalizations of CCCs and NCCCs. 
It is important
to distinguish between statistical, systematic, and intrinsic scatter and to take
the intrinsic scatter into account in applications (e.g. Ikebe et al.\ 2002;
Stanek et al.\ 2006). The findings here indicate that a significant portion
of the scatter may be intrinsic due to variations in the X-ray luminosity 
for clusters of a given mass. We are currently working to
confirm this result with high quality Chandra and XMM-Newton
observations of the complete HIFLUGCS sample. If a substantial
scatter is confirmed, it will be interesting to check
if these results are fully consistent with the low
$\Omega_{\rm m}$ and $\sigma_8$ values found from
the WMAP 3rd year data (Spergel et al. 2006) combined with
cluster mass-function prediction and observed luminosity function 
(Reiprich 2006). This finding also
points the way to an improvement in the $L_X - M_{500}$ relation. 
Given a good proxy
for the strength of cooling cores, one could correct for this effect and tighten
this important relation. This was indeed suggested by O'Hara et al. (2006) who
propose to use the central surface brightness as such a proxy. Further work is
in progress to use the HIFLUGCS cluster sample to work out a correction scheme.

\begin{acknowledgements}
Y. Chen was supported by the exchange program between
the Max-Planck Society and the Chinese Academy of Sciences.
He thanks MPE for its very gracious hospitality.
This research is partially supported by the Scientific Research 
Foundation for the
Returned Overseas Chinese Scholars, State Education Ministry.
\end{acknowledgements}

\end{document}